\documentclass[prd,preprint,tightenlines,floatfix,showpacs,preprintnumbers,nofootinbib,eqsecnum]{revtex4}
  \usepackage{amssymb}
   \usepackage{amsmath}
    \usepackage{amsfonts}
     \usepackage{epsfig}
      \usepackage{bm} 
      \usepackage[dvipsnames,usenames]{color}

 \def\gsim{\mathrel{\rlap{\lower4pt\hbox{\hskip1pt$\sim$}}
 \raise1pt\hbox{$>$}}}

 \newcommand\beq{\begin{equation}}
 
 \newcommand\eeq{\end{equation}}
 \newcommand\beqn{\begin{eqnarray}}
 \newcommand\eeqn{\end{eqnarray}}

\def\GeV{\,\mbox{GeV}}

\def\lsim{\mathrel{\rlap{\lower4pt\hbox{\hskip1pt$\sim$}}
    \raise1pt\hbox{$<$}}}         
\def\gsim{\mathrel{\rlap{\lower4pt\hbox{\hskip1pt$\sim$}}
    \raise1pt\hbox{$>$}}}         

\def\GeV{\,\mbox{GeV}}

\def\beq{\begin{equation}}
\def\eeq{\end{equation}}

\def\beqy{\begin{eqnarray}}
\def\eeqy{\end{eqnarray}}

\begin{document}

\begin{flushright}
LU TP 17-08\\
April 2017
\end{flushright}

\title{Heavy flavor production in high-energy $pp$ collisions:\\
color dipole description}

\author{Victor P.~Goncalves$^{1}$}
\email{barros@ufpel.edu.br}

\author{Boris Kopeliovich$^{2}$}
\email{bzk@mpi-hd.mpg.de}

\author{Jan Nemchik$^{3}$}
\email{nemcik@saske.sk}

\author{Roman Pasechnik$^{4}$}
\email{Roman.Pasechnik@thep.lu.se}

\author{Irina Potashnikova$^{2}$}
\email{irina.potashnikova@usm.cl}

\affiliation{
{$^1$\sl
High and Medium Energy Group, Instituto de F\'{\i}sica e Matem\'atica, 
Universidade Federal de Pelotas, Pelotas, RS, 96010-900, Brazil
}\\
{$^2$\sl 
Departamento de F\'{\i}sica,
Universidad T\'ecnica Federico Santa Mar\'{\i}a;\\
Centro Cient\'ifico-Tecnol\'ogico de Valpara\'{\i}so,
Casilla 110-V, Valpara\'{\i}so, Chile
}\\
{$^3$\sl 
Czech Technical University in Prague, FNSPE, B\v rehov\'a 7, 11519 
Prague, Czech Republic; \\
Institute of Experimental Physics SAS, Watsonova 47, 04001 Ko\v 
sice, Slovakia
}\\
{$^4$\sl
Department of Astronomy and Theoretical Physics, Lund
University, SE-223 62 Lund, Sweden
}
}

\begin{abstract}
\vspace{0.5cm}
We present a detailed study of open heavy flavor production in high-energy $pp$ collisions at the LHC in the color dipole framework. 
The transverse momentum distributions of produced $b$-jets, accounting for the jet energy loss, as well as produced open charm 
$D$ and bottom $B$ mesons in distinct rapidity intervals relevant for LHC measurements are computed. The dipole model results
for the differential $b$-jet production cross section are compared to the recent ATLAS and CMS data while the results for $D$ and $B$ 
mesons production cross sections -- to the corresponding LHCb data. Several models for the phenomenological dipole cross section 
have been employed to estimate theoretical uncertainties of the dipole model predictions. We demonstrate that the primordial transverse 
momentum distribution of the projectile gluon significantly affects the meson spectra at low transverse momenta and contributes to 
the largest uncertainty of the dipole model predictions.
\end{abstract}


\maketitle

\section{Introduction}
\label{Sec:intro}

Heavy flavor production in high-energy hadron-hardon collisions serves as a prominent testing ground for various perturbative QCD (pQCD) approaches 
(for a thorough review of the existing methods and results, see e.g. Refs.~\cite{Frixione:1997ma,Baines:2006uw,Andronic:2015wma}). During the last decade 
the experimental accuracy of heavy flavor production measurements in high-energy $pp$ collisions has been drammatically increased due to largely improved 
statistics and detection techniques at the LHC. Theoretical developments are expected to follow this trend by offering theoretical tools capable
to reproduce the observed energy dependence as well as transverse momentum and rapidity correlations for produced heavy quarks. This provides
a good baseline also for further predictions in various kinematic regions of future measurements.

One of such well-known and widely used tools is the QCD collinear factorisation approach \cite{Collins:1988ig,Collins:1989gx} which has been developed for 
heavy quark production up to the next-to-leading order (NLO) since a long time ago (see e.g. Refs.~\cite{Nason:1987xz,Altarelli:1988qr,Beenakker:1988bq,
Nason:1989zy,Beenakker:1990maa,levin91,levin92}). In collinear factorisation all the incident particles are assumed to be on-mass-shell carrying only 
longitudinal momenta, while the cross section is averaged over transverse polarisations of the incoming gluons. In this case, virtualities of the initial 
partons are taken into account only through the scale dependence of the corresponding structure functions, collinear parton distribution functions (PDFs), 
governed by the Dokshitzer-Gribov-Lipatov-Altarelli-Parisi (DGLAP) evolution equation \cite{Gribov:1972ri,Altarelli:1977zs,Dokshitzer:1977sg}. 
There are several popular approaches which attempt to resum large perturbative terms containing powers of $\alpha_s\log (p_T/m_Q)$ (e.g. leading-log 
$\alpha_s^n\log^n(p_T/m_Q)$ and next-to-leading log $\alpha_s^n\log^{n-1}(p_T/m_Q)$), where $p_T$ and $m_Q$ are the heavy quark 
transverse momentum and mass, respectively (see e.g. Refs.~\cite{Collins:1986mp,Olness:1987ep,Aivazis:1993kh,
Aivazis:1993pi,Buza:1995ie,Buza:1996wv,Martin:1996eva,Thorne:1997ga,Thorne:1997uu,Cacciari:1998it,Collins:1998rz,Kniehl:2004fy,
Kniehl:2005mk}). These approaches differ by the perturbative order at which the initial condition for a collinear PDF 
or a fragmentation function is computed and by the procedure of matching of resummed soft/collinear emissions with the fixed-order matrix elements.
In spite of the continuous progress over the last thirty years, the collinear factorisation approach suffers from such ambiguities as yet unknown 
higher-order process-dependent QCD corrections and scale (energy) dependence of the observables, as well as QCD factorisation breaking and 
medium-induced (such as saturation and energy loss) effects which are especially pronounced in heavy-ion collisions \cite{Andronic:2015wma}.

The formalism which incorporates the incident parton transverse momenta (or virtualities) in the center-of-mass frame of colliding nucleons is typically 
referred to as the $k_T$-factorisation approach \cite{gribov83,marchesini88,levin90,cata90,cata91,collins91}. In this approach,
the hard scattering matrix elements at small-$x$, for example, are computed by taking into account the virtualities and polarisation states of the incident 
gluons whose densities at a given transverse momentum $k_T$, momentum fraction $x$ and factorisation scale $\mu^2$ are controlled by the so-called 
unintegrated gluon distribution functions (UGDFs). In the $k_T$-factorisation approach, a major part of higher-order QCD corrections (in particular, due 
to initial-state radiation off the fusing partons) is effectively taken into account by means of the transverse momentum $k_T$ evolution of unintegrated PDFs. 
The latter carry a more detailed information about the structure of the incident nucleons than collinear gluon PDFs. Heavy quark production has been studied
in the framework of $k_T$-factorisation approach e.g. in Refs.~\cite{ryskin99,hagler,ryskin01,shab2004,saleev,Chachamis:2015ona}. Depending on a process
and kinematical regions concerned, $k_T$-factorisation is not a generic phenomenon and can be broken (see e.g. Ref.~\cite{Collins:2007nk,Rogers:2010dm,
DelDuca:2011ae}), e.g. by soft spectator interactions and the corresponding factorisation breaking effects are difficult to quantify.

The heavy quark production, especially at large $x_F$, can be successfully described within the color dipole framework which does not rely on 
QCD factorisation \cite{Kopeliovich:2005ym}. In particular, production of heavy flavor and quarkonia in $pp$ and $pA$ collisions in the dipole picture has been extensively 
studied in Refs.~\cite{Nikolaev:1994de,Nikolaev:1995ty,Kopeliovich:2001ee,Kopeliovich:2002yv}. The present work is a natural continuation of previous studies with 
the main objective to extend the dipole description to the $p_T$-dependent cross section, and to confront the results of the dipole approach with recent LHC data on 
heavy flavoured jets and mesons, in particular, open charm and beauty in various regions of the phase space.

The paper is organised as follows. In Section~\ref{Sec:GG-QQ}, a theoretical basis for heavy quark pair $Q\bar Q$ production in the dipole picture is presented and 
the production amplitudes derived. Section~\ref{Sec:dip-F} is devoted to derivation of the dipole formula for fully differential cross section of open heavy 
flavor production in both impact parameter and momentum representations. In Section~\ref{Sec:E-loss}, we briefly discuss the effect of leakage of energy from 
a jet cone of a restricted size, which leads to an effective shift of the jet transverse momentum. In Section~\ref{Sec:dip-par}, a few relevant parameterisations for the dipole 
cross section as the main phenomenological ingredient of the dipole formula have been reviewed. Section~\ref{Sec:results} presents numerical results for 
typical differential observables for open charm and bottom mesons, as well as for $b$-jets in comparison with recent ATLAS, CMS and LHCb data. Finally, 
in Section~\ref{Sec:summary} a short summary of our analysis is given.

\section{Heavy quark pair production in the dipole picture}
\label{Sec:GG-QQ}

It has been shown in several studies so far that in hard processes the dipole formalism effectively accounts for the higher-order QCD corrections and enables us 
to quantify such phenomena as the gluon shadowing, saturation, initial state interaction as well as nuclear coherence effects in a universal way (see e.g. 
Refs.~\cite{Kopeliovich:1981pz,Nikolaev:1993th,Nikolaev:1994kk,Nikolaev:1994de,Nikolaev:1995ty,Kopeliovich:1995an,Brodsky:1996nj,
Kopeliovich:1998nw,Kopeliovich:2000fb}. At small Bjorken $x$, the dipole formalism operates in terms of the eigenstates of interaction  \cite{Kopeliovich:1981pz}, namely, 
color dipoles with a definite transverse separation propagating through a color field of the target nucleon. In practice, this suggests to decompose 
any hadron-target scattering amplitude in the target rest frame into a superposition of universal ingredients -- the partial dipole-target scattering amplitudes 
$f_{\rm el}({\bf b},{\bf r};x)$ at different dipole separations ${\bf r}$ and impact parameters ${\bf b}$ convoluted with the light-cone distribution amplitudes 
for a given Fock state. 

In particular, the Deep-Inelastic Scattering (DIS) process at large $Q^2$ and small Bjorken $x$ in the target rest frame is viewed as a scattering of the ``frozen'' 
$q\bar q$ dipole of size $r\sim 1/Q$, originating as a fluctuation of the virtual photon $\gamma^* \to q\bar q$ with the 4-momentum squared $q^2=-Q^2$, off the target nucleon. 
The Drell-Yan (DY) pair production is considered in the target rest frame as a bremsstrahlung of massive $\gamma^*$ (and $Z^0$ boson) by the projectile quark before 
and after the quark scatters off the target, and thus can be viewed as a dipole-target scattering too \cite{Kopeliovich:1995an}. The projectile high-energy $q\bar q$ 
dipole probes the dense gluonic field in the target at high energies, when the nonlinear (e.g. saturation) effects due to multiple soft gluon interactions become relevant. 
Integrating the partial dipole amplitude over ${\bf b}$ one obtains the universal dipole cross section $\sigma_{\bar qq}=\sigma_{\bar qq}(r,x)$ that cannot be fully 
predicted from the first principles of perturbative QCD. Due to universality, however, this object is normally determined phenomenologically by fitting to e.g. DIS data 
(for more details, see Sect.~\ref{Sec:dip-par} below), and then such parameterisations can be used for description of all other sets of data on both inclusive and 
diffractive processes in $ep$, $pp$, $pA$ and $AA$ collisions.

\begin{figure*}[!h]
 \centerline{\includegraphics[width=0.85\textwidth]{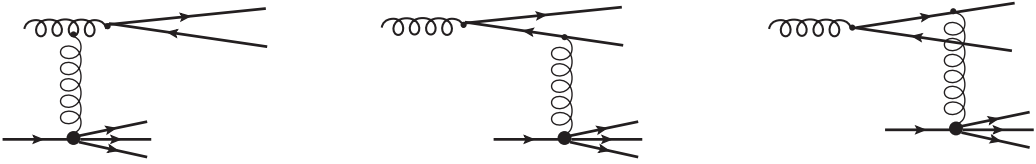}}
   \caption{
   \small Typical contributions to the heavy quark $Q\bar Q$ pair 
   production in $G_a\to Q\bar Q$ splitting subprocess in the color 
   field of the target nucleon.}
 \label{fig:GG-QQ}
\end{figure*}

Let us consider the color dipole formulation for inclusive production of a heavy quark pair and start with the leading-order process in gluon-proton scattering, 
\begin{eqnarray}
G_a + p \to Q\bar Q + X \,, \qquad Q=c,b \,.
\label{20}
\end{eqnarray}
In the target proton rest frame, the projectile gluon fluctuates into a $Q\bar Q$ pair as its relevant lowest-order Fock component, i.e. $G_a \to Q\bar Q$. 
The cross section can be presented as interaction of a  colorless 3-body system $G_aQ\bar Q$ scattering off the color background field of the target proton 
\cite{Nikolaev:1994de,Nikolaev:1995ty,Kopeliovich:2002yv}, as is illustrated in Fig.~\ref{fig:GG-QQ}. Here, $G_a$ is the initial gluon in a color 
state $a$, whose probability distribution over the fractional light-cone momentum $x_1$ is characterised by the gluon PDF in the incident hadron. Then in the dipole 
framework such leading-order contributions, after squaring and generalising to all orders, give rise to the dipole formula for the differential cross section written in terms
of the universal dipole cross section. This approach provides an effective way to incorporate real corrections due to the unresolved initial- and final-state radiation off 
the target gluon and the $Q\bar Q$ pair, respectively \cite{Nikolaev:1994de,Nikolaev:1995ty}.

The amplitude for inclusive $G_a + p \to Q\bar Q+X$ production in gluon-target scattering is then given by 
the sum of three contributions as is depicted in Fig.~\ref{fig:GG-QQ}, namely,
\begin{eqnarray}
\nonumber
A^{\mu\bar\mu}_{a}(\vec s, \vec r) &=&2\sqrt{3}\,\sum_{d=1}^{N_c^2-1} 
{\xi_Q^\mu}^\dagger  \Big\{\tau_d\,\tau_a\, \hat{\gamma}^{(d)}(\vec s + \bar\alpha \vec r) - 
\tau_a\,\tau_d\, \hat{\gamma}^{(d)}(\vec s - \alpha \vec r) \\ 
 &-& i\sum_{c} f_{cda} \tau_c \, \hat{\gamma}^{(d)}(\vec s) \Big\}
\hat \Phi_{Q\bar Q}(\alpha,\vec r)\,\tilde{\xi}_{\bar Q}^{\bar \mu} \,, \qquad  
\tilde{\xi}_{\bar Q}^{\bar\mu}=i\sigma_y (\xi_{\bar Q}^{\bar\mu})^* \,, \label{QQ-amp}
\end{eqnarray} 
where $\hat \gamma^{(a)}(\vec{s})$ is the gluon-target interaction amplitude, $\alpha$ ($\bar\alpha=1-\alpha$) 
is the light-cone momentum fraction of the gluon carried by the heavy quark (antiquark), $\tau_a$ are 
the standard $SU(N_c)$ generators related to the Gell-Mann matrices as $\lambda_a=\tau_a/2$, 
$\vec{s}$ is the transverse distance between projectile gluon and the center of gravity of the target,
$\hat \Phi_{Q\bar Q}$ is the distribution amplitude of the $G_a\to Q\bar Q$ splitting,
and $\xi_Q^\mu$ are the 2-spinors normalised as
\begin{eqnarray}
\sum_{\mu,\bar\mu}\tilde{\xi}_{\bar Q}^{\bar \mu}
\big({\xi_Q^\mu}^\dagger\big)^*= \hat{1} \,, \qquad
\sum_{\mu,\bar\mu}\big({\xi_Q^\mu}^\dagger \hat a  \tilde{\xi}_{\bar Q}^{\bar \mu} \big)^*\, 
\big({\xi_Q^\mu}^\dagger \hat b  \tilde{\xi}_{\bar Q}^{\bar \mu}\big)  = {\rm Tr}\big(\hat a^\dagger \hat b\big) \,.
\end{eqnarray}
The amplitude $\hat \Phi_{Q\bar Q}$ in impact parameter representation is given by
\begin{eqnarray} 
\label{QQ-wf}
\hat \Phi_{Q\bar Q}(\alpha,\vec r) = \frac{\sqrt{\alpha_s}}{(2\pi)\sqrt{2}} \, 
\Big\{m_Q(\vec e\cdot \vec \sigma)+i(1-2\alpha)(\vec\sigma\cdot \vec n)(\vec e\cdot \vec\nabla_r)-
(\vec e\times \vec n)\cdot \vec\nabla_r \Big\}\,K_0(m_Q\, r) \,,
\end{eqnarray}
where $\alpha_s$ is the QCD coupling, $\vec n$ is the unit vector parallel to the gluon momentum, 
$\vec e$ is the polarisation vector of the gluon, $\vec\sigma$ is the 3-vector of the Pauli spin-matrices, 
$K_0(x)$ is the modified Bessel function of the second kind, 
and $\vec{\nabla}_r\equiv \partial/\partial \vec r$. 

Following Ref.~\cite{Kopeliovich:2002yv}, the total inclusive $Q\bar Q$ production amplitude can be
separated into a superposition of color-singlet and color-octet contributions which are odd and 
even under permutation of non-color variables (spatial and spin indices) of the $Q$ and 
$\bar Q$ quarks as follows
\begin{eqnarray} 
\label{ampl}
A^{\mu\bar\mu}_{a}={\xi_Q^\mu}^\dagger\Big\{ A_{a,1^-}+A_{a,8^-}+A_{a,8^+}\Big\}\,\tilde{\xi}_{\bar Q}^{\bar \mu}\,, 
\end{eqnarray}
where
\begin{eqnarray}
&& A_{a,1^-}(\vec s,\vec r)=\frac{1}{6}\sum_d\delta_{ad}\delta_{ij}\,O^{(d)}(\vec s,\vec r) \,, \\ 
&& A_{a,8^-}= \frac{1}{2}\sum_{d,g}d_{adg}(\tau_g)^i_j\,O^{(d)}(\vec s,\vec r)\,, \\ 
&& A_{a,8^+}= \frac{i}{2}\sum_{d,g}f_{adg}(\tau_g)^i_j\,E^{(d)}(\vec s,\vec r)\,, 
\end{eqnarray}
and $f_{abc}$ and $d_{abc}$ ($a,b,c=1,\dots,8$) are the antisymmetric and symmetric $SU(3)$ structure constants, respectively.
Then negative parity with respect to such an interchange corresponds to $Q\bar Q$ state with positive $C$-parity ($C$-even) 
and is denoted as $1^-$ for color singlet and $8^-$ for color octet, and vice versa. Here, the odd $O$ and even $E$ factors read
\begin{eqnarray}
&& O^{(d)}(\vec s,\vec r)=2\sqrt{3}\,\hat \Phi_{Q\bar Q}(\alpha,\vec r)\,
\Big[ \hat{\gamma}^{(d)}(\vec s - \alpha \vec r) - \hat{\gamma}^{(d)}(\vec s + \bar\alpha \vec r) \Big] \,, \\
&& E^{(d)}(\vec s,\vec r)=2\sqrt{3}\,\hat \Phi_{Q\bar Q}(\alpha,\vec r)\,
\Big[ \hat{\gamma}^{(d)}(\vec s - \alpha \vec r) + \hat{\gamma}^{(d)}(\vec s + \bar\alpha \vec r) - 
2\hat{\gamma}^{(d)}(\vec s)\Big] \,,
\end{eqnarray}
respectively.

When taking square of the total inclusive $G_a+p\to Q\bar Q+X$ amplitude\footnote{Note, averaging over the projectile gluon polarisation 
$\lambda_*$ is normally accounted for in the normalisation of the corresponding $G\to Q\bar Q$ wave function (\ref{QQ-wf}), by convention.}
\begin{eqnarray}
\overline{ |A|^2 }(\vec r_1;\vec r_2)\equiv \frac{1}{8}\,\int d^2 s\, d\{X\}\sum_{\lambda_{*},a,\mu,\bar\mu}
\Big\langle A^{\mu\bar\mu}_{a}(\vec s, \vec r_1) 
\big(A^{\mu\bar\mu}_{a}\big)^\dagger(\vec s, \vec r_2)\Big\rangle \label{A2r1r2}
\end{eqnarray}
one performs an averaging over color indices $a$ and, implicitly, over polarisation $\lambda_*$ 
of the incoming projectile gluon $G_a$ as well as valence quarks and their relative coordinates 
in the target nucleon. By the optical theorem, the universal dipole cross section $\sigma_{\bar qq}(\vec \rho)$
is related to the partial dipole elastic amplitude $\mathrm{Im}f_{el}(\vec s,\vec \rho)$, which is given in terms of 
the square of inelastic scattering amplitude
\begin{eqnarray} \label{G-N}
\hat{C}^{(d)}(\vec s,\vec \rho)\equiv \gamma^{(d)}(\vec s) - \gamma^{(d)}(\vec s + \vec\rho) \,,
\end{eqnarray}
as follows
\begin{eqnarray}
\int d^2s \sum_X\langle i | \hat{C}^{(d)}(\vec s,\vec \rho)\hat{C}^{(d')}(\vec s,\vec \rho) | i \rangle \equiv
\frac18\, \delta_{dd'}\,\int d^2s\; 2\mathrm{Im}f_{el}(\vec s,\vec \rho)=\frac18\, \delta_{dd'}\,\sigma_{\bar qq}(\vec \rho) \,. 
\label{C2}
\end{eqnarray}
This relation can be used in practical derivations of the dipole formula for 
differential cross sections. 

The dipole cross section is related to the intrinsic dipole transverse momentum distribution (dipole TMD in what follows) 
${\cal K}_{\rm dip}(x,\kappa_\perp^2)$ as \cite{Nikolaev:1990ja,bgbk}
\begin{eqnarray} \label{dipole-UGDF}
\sigma_{\bar qq}(\vec r,x)\equiv \frac{2\pi}{3}\int \frac{d^2\kappa_\perp}{\kappa_\perp^4}\,
(1-e^{i\vec \kappa_\perp\cdot \vec r})(1-e^{-i\vec \kappa_\perp\cdot \vec r})\,{\cal K}_{\rm dip}(x,\kappa_\perp^2) \,.
\end{eqnarray}
In the perturbative QCD language, at sufficiently large target gluon transverse momentum $\kappa_\perp \gg \Lambda_{\rm QCD}$ 
the dipole TMD is approximately equal to the unintegrated gluon distribution function times $\alpha_s$ pointing at a connection 
between the $k_\perp$-factorisation and dipole approaches. Indeed, in the double logarithmic approximation
of the DGLAP equations, one has the following relation at large $Q^2$ \cite{bgbk}
\begin{eqnarray}
\frac{1}{\pi}\int^{Q^2} \frac{d^2\kappa_\perp}{\kappa_\perp^2}\,{\cal K}_{\rm dip}(x,\kappa_\perp^2)=\alpha_s(Q^2)xg(x,Q^2) \,.
\end{eqnarray}
Such a relation between the dipole TMD ${\cal K}_{\rm dip}(x,\vec \kappa_\perp^2)$, extracted from a known model for $\sigma_{\bar qq}$, 
and conventional UGDF ${\cal F}(x,\vec \kappa_\perp^2)$ is, however, only approximate and does not hold e.g. in the soft $\kappa_\perp$ region 
(as well as at large $x$) corresponding to large $q\bar q$ dipole separations where the saturation is effective and the conventional UGDF 
is not well defined, so the dipole cross section $\sigma_{\bar qq}$ should be used. In the dipole framework we go beyond $k_T$-factorisation where 
${\cal F}(x,\vec \kappa_\perp^2)$ represents a two-gluon amplitude. In this case, for any $\kappa_{\perp}$ one could employ 
Eq.~(\ref{dipole-UGDF}) as a formal definition of the dipole TMD built upon a known parameterisation of the universal dipole cross 
section following Ref.~\cite{bgbk}, i.e. 
\begin{eqnarray} \label{dipole-UGDF-1}
\frac{1}{\kappa_\perp^4}{\cal K}_{\rm dip}(x,\kappa_\perp^2)=\frac{3}{8\pi^2}\int^\infty_0 dr\,r\,J_0(\kappa_\perp r)\, 
\Big[ \sigma_{\bar qq}^\infty(x) - \sigma_{\bar qq}(r,x) \Big] \,,
\end{eqnarray}
where $\sigma_{\bar qq}^\infty(x)={\rm lim}_{r\to\infty}\sigma_{\bar qq}(r,x)$ and $J_0(x)$ is the Bessel function of the first kind.

Production of heavy ($c$ and $b$) quark pairs is associated with dipoles of small size, so that the use of the approximate relation
\begin{eqnarray} 
\label{dip-vs-kt}
{\cal K}_{\rm dip}(x,\kappa_\perp^2) \simeq \alpha_s\,{\cal F}(x,\kappa_\perp^2)
\end{eqnarray}
is justified in the whole experimentally accessible range of the heavy quark transverse momenta. One of our particular goals is to test the relation (\ref{dip-vs-kt}) 
and an impact of possible deviations from it depending on heavy quark $p_T$ and $y$ spectra measured at the LHC. As we will see below, the dipole TMD is a highly convenient 
object in practical calculations within the dipole approach enabling to formulate the dipole formula for heavy quark pair production explicitly in momentum representation.

\section{Dipole formula for open heavy flavor production}
\label{Sec:dip-F}

Following to the above scheme one can obtain the amplitude squared $\overline{|A|^2}$
in an analytic form as a linear combination of the dipole cross sections for different dipole separations,
with coefficients given by the color structure and distribution amplitudes of the considering Fock state. 
As a starting point, the cross section differential in the (anti)quark $p_T$ and momentum 
fraction $\alpha$ corresponding to the $G_a+p \to Q + X$ process is given by
\begin{eqnarray} 
\frac{d\sigma_{Gp\to QX}}{d\alpha\, d^2 p_T} = \frac{1}{(2\pi)^2}\,
\int d^2r_1d^2r_2\; e^{i\vec p_T \cdot (\vec r_1- \vec r_2)}\, \overline{|A|^2}(\vec r_1;\vec r_2) \,,
\label{Gp-CS}
\end{eqnarray}
where the amplitude squared can be found by means of Eqs.~(\ref{ampl}), (\ref{A2r1r2}) and (\ref{C2}). 
Integrating over $p_T$ and $\alpha$, one arrives at the dipole formula for the total cross section
\begin{eqnarray}
\sigma_{Gp \to QX} = \int d\alpha \int d^2r\; |\Phi_{Q\bar Q}(\alpha,\vec r)|^2 \sigma_{q\bar qG}(\alpha,\vec r)\,,
\end{eqnarray}
where $\sigma_{q\bar qG}$ is the effective 3-body dipole cross section can be expressed as a sum of $Q\bar Q$ 
singlet $1^-$ and octet $8^\pm$ contributions,
\begin{eqnarray}
\sigma_{q\bar qG}(\alpha,\vec r) \equiv \sum_{S=1^-,8^\pm}\sigma^S_3 = \frac{9}{8}\Big(\sigma_{q \bar q}(\bar\alpha \vec r) + 
\sigma_{q \bar q}(\alpha \vec r)  \Big) - \frac{1}{8} \sigma_{q \bar q}(\vec r) \,, \label{sig3}
\end{eqnarray}
where
\begin{eqnarray}
\sigma^{1^-}_3 = \frac{1}{8}\sigma_{q \bar q}(\vec r)\,, \quad 
\sigma^{8^-}_3 = \frac{5}{16}\sigma_{q \bar q}(\vec r)\,, \quad 
\sigma^{8^+}_3 = \frac{9}{16}\Big[2\sigma_{q \bar q}(\alpha \vec r) + 
2\sigma_{q \bar q}(\bar\alpha \vec r) - \sigma_{q \bar q}(\vec r) \Big]\,. \nonumber
\end{eqnarray}
The $G\to Q\bar Q$ transition amplitude squared reads \cite{Nikolaev:1994de,Nikolaev:1995ty}
\begin{eqnarray} \nonumber
|\Phi_{Q\bar Q}(\alpha,\vec r)|^2 &\equiv& \sum_{\lambda_*=\pm 1}
\mathrm{Tr}\Big[\hat \Phi_{\bar QQ}(\alpha,\vec r)\cdot \hat \Phi^\dagger_{\bar QQ}(\alpha,\vec r) \Big] \\ &=&
\frac{\alpha_s}{(2\pi)^2}\,\Big[m_Q^2 K_0^2(m_Q\, r) + (\alpha^2 + \bar \alpha^2) m_Q^2 K_1^2(m_Q\, r)  \Big] \,,
\end{eqnarray}
where
\[
\frac{\vec{r}}{r} K_1(r)=-\vec{\nabla}_r K_0(r) \,.
\]

In what follows, we are interested in analysis of the inclusive (in color and parity) cross section, differential in $p_T$,
which can be written as,
 \begin{eqnarray}
\frac{d^3\sigma_{Gp \to QX}}{d\alpha\, d^2 p_T } = \frac{1}{(2\pi)^2} \int d^2r_1 d^2r_2 \;
e^{i \vec{p}_T\cdot (\vec{r}_1 - \vec{r}_2)}\, \Psi^*_{Q \bar{Q}} (\alpha, \vec{r}_1) 
\Psi_{Q \bar{Q}} (\alpha, \vec{r}_2) \sigma_{\rm eff} (\alpha,\vec{r}_1, \vec{r}_2) \,,
\label{spectra:ip}
\end{eqnarray} 
where
\begin{eqnarray}
\Psi^*_{Q \bar{Q}} (\alpha, \vec{r}_1) \Psi_{Q \bar{Q}} (\alpha, \vec{r}_2) & = & 
\frac{\alpha_s}{(2\pi)^2} \left[m_Q^2 K_0(m_Q r_1) K_0(m_Q r_2) \right. \nonumber \\
& \, & \left. +  (\alpha^2 + \bar \alpha^2) m_Q^2 \frac{\vec{r}_1\cdot \vec{r}_2}{r_1 r_2} K_1(m_Q\, r_1)K_1(m_Q\, r_2)\right] \,.
\label{wf}
\end{eqnarray}
The effective dipole cross section is given by
\begin{eqnarray}
\sigma_{\rm eff} (\alpha,\vec{r}_1, \vec{r}_2) & = &\frac{9}{16} \sigma_{q\bar{q}} (\alpha \vec{r}_1)  + 
\frac{9}{16} \sigma_{q\bar{q}} (\bar{\alpha} \vec{r}_1) + \frac{9}{16} \sigma_{q\bar{q}} (\alpha \vec{r}_2) + 
\frac{9}{16} \sigma_{q\bar{q}} (\bar{\alpha} \vec{r}_2) \nonumber \\
& - & \frac{1}{16} \sigma_{q\bar{q}} (\bar{\alpha} \vec{r}_1 + {\alpha} \vec{r}_2) - 
\frac{1}{16} \sigma_{q\bar{q}} ({\alpha} \vec{r}_1 + \bar{\alpha} \vec{r}_2) \nonumber \\
& - & \frac{1}{2} \sigma_{q\bar{q}} ({\alpha} |\vec{r}_1 - \vec{r}_2|) - 
\frac{1}{2}\sigma_{q\bar{q}} (\bar{\alpha} |\vec{r}_1 - \vec{r}_2|) \,.
\label{sig-eff}
\end{eqnarray}

In general, a transition from $Gp$ to $pp$ scattering implies that the projectile gluon is not collinear any more but can carry 
a transverse momentum relative to the beam proton. The resulting $pp\to QX$ cross section can be obtained by 
an appropriate shift of kinematic variables and by a convolution of the $Gp \to QX$ cross section with the projectile 
gluon UGDF similarly to that in the $k_T$-factorisation approach. Taking into account the transverse momentum $k_T$ of 
the incident gluon, the $pp\to QX$ cross section then reads,
\begin{eqnarray}
\frac{d\sigma_{pp \rightarrow QX}}{dyd\alpha d^2p_T} = \int \frac{d^2 k_T}{k_T^2}\,d^2p_T^\prime\,{\cal F}(x_1,k_T^2) \,
\frac{d\sigma_{Gp \to  QX}}{d\alpha d^2p_T^{\prime}}\,
\delta\left(\vec p_T^{\,\prime}-\vec p_T+\alpha \vec k_T\right)\,,
\label{factkt}
\end{eqnarray}
where ${\cal F}(x_1,k_T^2)$ is the unintegrated gluon distribution of the incident gluon with momentum fraction $x_1$. 
If the primordial gluon momentum were disregarded, one would obtain 
\begin{eqnarray} 
 \frac{d\sigma_{pp \rightarrow QX}}{dy d\alpha d^2p_T} = G(x_1,\mu^2)\,\frac{d\sigma_{Gp \to QX}}{d\alpha d^2p_T} \,, 
 \label{dipole-f}
\end{eqnarray}
where the projectile gluon distribution in the incoming proton,
\begin{equation} 
 G(x_1,\mu^2)\equiv x_1g(x_1,\mu^2) = \frac{1}{\pi}\int^{\mu^2} \frac{d^2 k_T}{k_T^2}\,{\cal F}(x_1,k_T^2) \,.
\label{g}
\end{equation}

All dipole cross sections, introduced above, implicitly depend on target fractional light-cone momentum $x_2$. The values of $x_1$ and $x_2$ 
can be estimated in the LO process $G_1+G_2\rightarrow\bar QQ$ in the collinear approximation,
\begin{equation} 
x_{1,2} = \frac{M_{Q\bar Q}}{\sqrt{s}}\, e^{\pm y}\,, \quad M_{Q\bar Q}\simeq 2\sqrt{m_Q^2+p_T^2}\,.
\label{x1x2}
\end{equation}
We also use the invariant mass of the $Q\bar Q$ pair $M_{Q\bar Q}$, as the scale $\mu^2=M_{Q\bar Q}^2$ in Eq.~(\ref{g}) and further calculations.

The analysis of heavy quark $p_T$ spectra in the impact parameter space using Eq.~(\ref{spectra:ip}) implies 
the calculation of 2-dim Fourier integrals of products of Bessel functions and the dipole cross 
section which is numerically challenging for a generic dipole parameterisation.
On the other hand, starting from the dipole formula in impact parameter representation (\ref{spectra:ip}), 
the relation (\ref{dipole-UGDF-1}) enables us to obtain a much simpler expression for the heavy quark 
$p_T$ spectrum manifestly in momentum representation
\begin{eqnarray} \nonumber
\frac{d^3\sigma_{Gp \to QX}}{d\alpha d^2 p_T } &=& 
\frac{1}{ 6 \pi}  \int \frac{d^2 \kappa_\perp}{\kappa_\perp^4}  \alpha_s(\mu^2)\, {\cal K}_{\rm dip}(x,\kappa_\perp^2)\,
\Big\{\Big[\frac{9}{8}{\cal{H}}_0(\alpha,\bar{\alpha},p_T) - \frac{9}{4} {\cal{H}}_1(\alpha,\bar{\alpha},\vec{p}_T,\vec{\kappa}_\perp) \\ 
&+& {\cal{H}}_2(\alpha,\bar{\alpha},\vec{p}_T,\vec{\kappa}_\perp) + \frac{1}{8}{\cal{H}}_3(\alpha,\bar{\alpha},\vec{p}_T,\vec{\kappa}_\perp)\Big] + 
\left[ \alpha \longleftrightarrow \bar{\alpha}\right]\Big\} \,,  \label{spectra:ms} 
\end{eqnarray}
where the dipole TMD ${\cal K}_{\rm dip}$ is defined by means of a known dipole cross section parameterisation (\ref{dipole-UGDF}), and
\begin{eqnarray} \nonumber
{\cal{H}}_0(\alpha,\bar{\alpha},p_T) & = & \frac{m_Q^2 + (\alpha^2 + \bar \alpha^2)p_T^2}{(p_T^2 + m_Q^2)^2} \,\,,\\
{\cal{H}}_1(\alpha,\bar{\alpha},\vec{p}_T,\vec{\kappa}_\perp)& = & \frac{m_Q^2 + (\alpha^2 + \bar \alpha^2) \vec{p}_T\cdot 
(\vec{p}_T - \alpha \vec{\kappa}_\perp ) }{[(\vec{p}_T - \alpha \vec{\kappa}_\perp)^2 + m_Q^2](p_T^2 + m_Q^2)}
\,\,, \nonumber \\
{\cal{H}}_2(\alpha,\bar{\alpha},\vec{p}_T,\vec{\kappa}_\perp)& = & \frac{m_Q^2 + (\alpha^2 + \bar \alpha^2) 
(\vec{p}_T - \alpha \vec{\kappa}_\perp )^2 }{[(\vec{p}_T - \alpha \vec{\kappa}_\perp)^2 + m_Q^2]^2}
\,\,, \nonumber \\
{\cal{H}}_3(\alpha,\bar{\alpha},\vec{p}_T,\vec{\kappa}_\perp)& = & \frac{m_Q^2 + (\alpha^2 + \bar \alpha^2) 
(\vec{p}_T + \alpha \vec{\kappa}_\perp )\cdot (\vec{p}_T - \bar{\alpha} \vec{\kappa}_\perp ) }
{[(\vec{p}_T + \alpha \vec{\kappa}_\perp)^2 + m_Q^2][(\vec{p}_T - \bar{\alpha} \vec{\kappa}_\perp)^2 + m_Q^2]}\,\,.
\end{eqnarray}

In the heavy quark limit the characteristic dipole sizes are small, so one can disregard the saturation behavior of the generic dipole parameterisation 
in Eq.~(\ref{sig-eff}) and rely on the small-$r$ approximation,
\begin{eqnarray}
\sigma_{\bar qq}(x,\vec r)  = C(x,\mu^2) \cdot {r}^2 \,, 
\label{CT}
\end{eqnarray}
where $C(x,\mu^2)$ is a model-dependent function of the target gluon fraction $x=x_2$ and, in general, the hard scale $\mu^2$. 
In this case, the 3-body effective dipole cross section $\sigma_{\rm eff} (\vec{r}_1, \vec{r}_2, \alpha)$ in Eq.~(\ref{sig-eff}) takes the simple form,
\begin{eqnarray}
\sigma_{\rm eff} (\vec{r}_1, \vec{r}_2, \alpha) \approx C(x_2,\mu^2) \cdot \left[\alpha^2 +  \bar{\alpha}^2 - 
\frac{1}{4} \right] \, \vec{r}_1 \cdot \vec{r}_2 \,,
\end{eqnarray}
which leads to the approximate result,
\begin{eqnarray} \nonumber
\frac{d^3\sigma_{G \to Q \bar Q}}{d\alpha d^2 p_T } = \frac{\alpha_s(\mu^2)\,C(x_2,\mu^2)}{(2\pi)^2} 
\left[\alpha^2 + \bar{\alpha}^2 - \frac{1}{4} \right] \left\{\frac{4 m_Q^2 p_T^2 }
{(m_Q^2 + p_T^2)^4} + (\alpha^2 + \bar{\alpha}^2) \frac{2 (m_Q^4 + p_T^4)}{(m_Q^2 + p_T^2)^4} \right\} \,, \\
\label{r2}
\end{eqnarray}
used further for the numerical analysis of inclusive high-$p_T$ $b$-jet production.

The differential distribution of open heavy flavored mesons ($M \equiv D,\,B$), produced in $pp$ collisions, can be found 
convoluting with the fragmentation function,
\begin{eqnarray} \label{mes-CS}
\frac{d\sigma_{pp \rightarrow M X}}{dY d^2P_T} = \int_{z_{\rm min}}^1 \frac{dz}{z^2} 
D_{Q/M} (z,\mu^2) \int_{\alpha_{\rm min}}^1 d\alpha \frac{d\sigma_{pp \rightarrow QX}}{dyd\alpha d^2p_T} \,,
\end{eqnarray}
where $z$ is the fractional light-cone momentum of the heavy quark carried by the meson $M$; 
$D_{Q/M}(z,\mu^2)$ is the fragmentation function; and
\begin{eqnarray}
\vec p_T = \vec P_T/z\,,\quad Y = y \,, \quad z_{\rm min}=\frac{\sqrt{m_M^2+P_T^2}}{\sqrt{s}}\,e^{Y} \,, \quad 
\alpha_{\rm min}=\frac{z_{\rm min}}{z}\sqrt{\frac{m_Q^2 z^2 + P_T^2}{m_M^2 + P_T^2}} \,, 
\end{eqnarray}
in terms of meson mass $m_M$, rapidity $Y$ and transverse momentum $P_T$. In the numerical calculations below, 
we rely on the DGLAP evolved parametrization of the fragmentation function $D_{Q/M}(z,\mu^2)$ fitted to LEP and SLAC data 
on $e^+e^-$ annihilation \cite{KKSS,BKK}. 

\section{Energy leakage off the jet}
\label{Sec:E-loss}

Detecting high-$p_T$ jets one avoids the necessity of convolution with the quark fragmentation function. This offers opportunity 
of direct comparison of the calculated $p_T$ dependent quark production cross section  with data. Besides, one can reach much higher 
values of $p_T$, which are strongly suppressed by the fragmentation function in the case of inclusive single hadron production. 
\begin{figure*}[!h]
 \centerline{\includegraphics[width=0.7\textwidth]{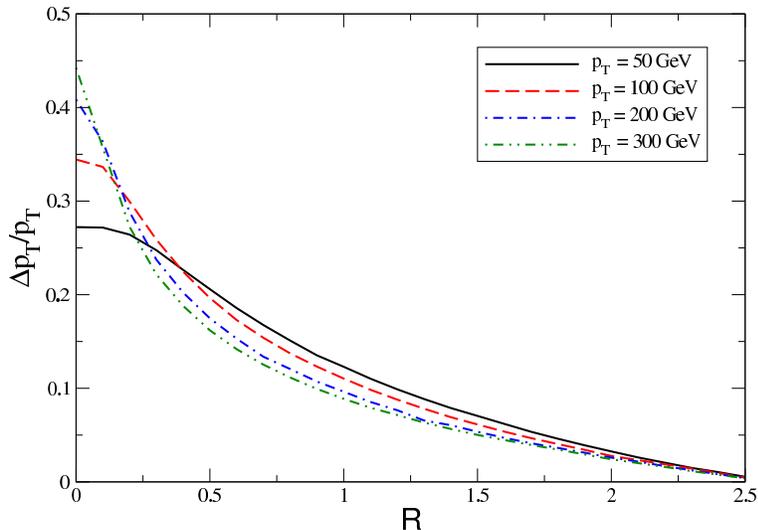}}
   \caption{ \small 
   (Color onine) The relative fraction of the jet transverse momentum $p_T$ radiated outside the measured jet 
   cone as a function of jet radius $R$ for various $p_T$ values.}
 \label{fig:jet-cone}
\end{figure*}

If the jet is detected within a cone $0<\theta<\theta_0$ relative to the jet axis, the fractional  momentum, radiated outside this cone is,
\beq
\frac{\Delta p_T}{p_T}={1\over v}\int\limits_{\lambda^2}^{p_T^2} dk^2 
\int\limits_{x_{min}}^1 dx\,x\,\frac{dn_g}{dxdk^2}\,
\Theta\left[\arctan\left(\frac{4p_T\,\tilde x\,\tilde k}{4p_T^2\,\tilde x^2-\tilde k^2}\right)-\theta_0\right].
\label{20}
\eeq
We use here the shorthand notations, $\tilde x=x(1+v)/2$; $\tilde k=kv$; and 
$v=p_T/\sqrt{p_T^2+m_Q^2}$. 
The bottom limit $\tilde x_{min}=\tilde k/p_T$.
The infrared cutoff $\lambda=0.65\,\rm GeV$ corresponds to the mean transverse momentum of gluons 
in the proton \cite{kst2,spots}, which can also be treated as an effective gluon mass.
The gluon radiation spectrum has the form \cite{gb},
\beq
\frac{dn_g}{dxdk^2}=
\frac{2\alpha_s(k^2)}{3\pi\,x}\,
\frac{k^2(2-2x+x^2)}{(k^2+x^2 m_Q^2)^2}.
\label{40}
\eeq
The running coupling  $\alpha_s(k^2)$ is taken in the one-loop approximation,
\beq
\alpha_s(k^2)=\frac{12\pi}{(33-2n_f)\ln[(k^2+k_0^2)/\Lambda_{\rm QCD}^2]}.
\label{60}
\eeq
Apparently, for radiation of a high-$p_T$ $b$-quark we should include all quarks up to $b$,
i.e. $n_f=5$. To regularize $\alpha_s(k^2)$ at low $k$ we modified the argument $k^2\to k^2+k_0^2$ with $k_0^2=0.5\GeV^2$.

Jet radius, defined as $R^2=\Delta\phi^2+\Delta\eta^2=8\theta_0^2$ controls the amount of energy radiated outside the cone. 
The relative variation of the jet transverse momentum, caused by this leakage of energy is depicted vs jet radius in Fig.~\ref{fig:jet-cone}.
As long as the jet $p_T$ and radius are known from a particular jet measurement, the relative fraction of 
the jet transverse momentum $\Delta p_T/p_T$ lost into radiation outside the cone angle $\theta_0$ can be found
from Fig.~\ref{fig:jet-cone}. In Section~\ref{Sec:results}, we analyse the energy leakage effect in numerical results for 
$p_T$ distributions of $b$ jets at the LHC.

\section{Dipole cross section and unintegrated gluon density}
\label{Sec:dip-par}
The universal dipole cross section first introduced in \cite{Kopeliovich:1981pz}, underwent  essential development, in particular its $x$-dependence, 
during last two decades, being strongly motivated by appearance of comprehensive experimental information from HERA.

A number of phenomenological models for the universal dipole cross section has become available in the literature during the last decade 
\cite{GBW,iim,kkt,dhj,Goncalves:2006yt,buw,kmw,agbs,Soyez2007,bgbk,kt,ipsatnewfit,amirs}). These parameterisations 
are conventionally based on saturation physics and in most cases rely on fits to the HERA data. One way to estimate theoretical uncertainties 
of the dipole model predictions is by comparing the numerical results obtained with distinct dipole parameterisations.

A saturated shape of the dipole cross section, first proposed in \cite{GBW} has the form,
\begin{equation}
\label{saturation}
\sigma_{q\bar{q}}(r,x) = \sigma_0\,
\left(1 - e^{-\frac{r^2\,Q_s^2(x,\mu^2)}{4}}\right)\,.
\end{equation}
reminding the Glauber model of multiple interactions, which also leads to saturation of nuclear effects. 
Correspondingly, the factor $C(x,\mu^2)$ introduced in Eq.~(\ref{CT}), has the form,
$C(x,\mu^2) =\sigma_0\,Q_s^2(x,\mu^2)/4$,
where $Q_s(x,\mu^2)$ is the saturation scale, which depends on $x=x_2$ and, in general, also on 
the hard scale $\mu^2=\mu^2(r)$, determined by a typical dipole separation $r=|\vec{r}|$. 
As long as $Q_s^2(x,\mu^2(r))$ is  a slow (e.g. logarithmic) function of the dipole 
separation $r$, one could employ the relation (\ref{dipole-UGDF-1}) such that the dipole TMD 
takes an approximate Gaussian shape
\begin{eqnarray}
\label{dugdf}
 {\cal K}_{\rm dip}(x,\kappa_\perp^2) \simeq  \frac{3 \sigma_0}{4 \pi^2}\frac{\kappa_\perp^4}{Q_s^2(x,\mu^2)} 
 e^{-\frac{\kappa_\perp^2}{Q_s^2(x,\mu^2)}}\,, \qquad \mu^2=\mu^2(\kappa_\perp)\,,
\end{eqnarray}
in terms of the main ingredients of the dipole cross sections, namely, its normalisation $\sigma_0$ and the saturation scale 
$Q_s^2(x,\mu^2)$, where the hard scale $\mu^2=M_{Q\bar Q}^2$. Note that the relation (\ref{dugdf}) 
is generic as long as the ansatz for the dipole cross section (\ref{saturation}) is imposed with $Q_s^2(x,\mu^2)$ being 
a slow function of $\mu^2$.

A simple and practical parameterisation of the saturation scale as function of $x$ and independent of $\mu^2$ was proposed in Ref.~\citep{GBW} 
(referred to as the Golec-Biernat--Wusthoff (GBW) model in what follows),
\begin{eqnarray}
\nonumber
\mathrm{GBW:}\qquad && Q_s^2 = Q_s^2(x) \equiv Q_0^2\left( \frac{x_0}{x} \right)^\lambda \,, \quad 
Q_0^2 = 1\,\mathrm{GeV}^2\,, \\
&& x_0 = 4.01 \times 10^{-5}\,, \quad \lambda = 0.277\,, \quad \sigma_0 = 29\, \mathrm{mb} \,, \label{gbw}
\end{eqnarray}
where parameters were extracted from fits of the saturated ansatz (\ref{saturation}) to the DIS HERA data accounting 
for a charm quark contribution. Such a naive phenomenological model has provided an overall good description of 
a wealth of experimental data on various production cross sections in hadronic collisions at small $x\lesssim 0.01$, 
which is effective at very high energies at the LHC, and for not very large momentum scales. 

In Refs.~\cite{Blaettel:1993rd,Frankfurt:1993it,Frankfurt:1996ri} it was understood that the dipole cross section at small 
separations $r$ can be related to the target gluon density as
\begin{eqnarray}
\sigma_{q\bar q}\simeq \frac{\pi^2}{3}\alpha_s\Big(\frac{\Lambda}{r^2}\Big)\,r^2\,xg\Big(x,\frac{\Lambda}{r^2}\Big)\,,
\end{eqnarray}
where $\Lambda\approx 10$ represents a numerical factor determined in Ref.~\cite{Nikolaev:1994cn}.
For dipole parameterisations including both the QCD DGLAP evolution of the target gluon density at the hard scale $\mu^2$ 
and the saturation, one of the first versions is proposed in Ref.~\cite{bgbk} and is denoted as the BGBK model 
in what follows. It uses the same saturated ansatz as in the GBW model (\ref{saturation}) but introduces an explicit collinear 
gluon PDF dependence into the saturation scale as follows
\begin{equation}
\label{bgbk}
\mathrm{BGBK:}\qquad Q_s^2 = Q_s^2(x,\mu^2) \equiv \frac{4\pi^2}{\sigma_0 N_c}\, \alpha_s(\mu^2)\,xg(x,\mu^2) \,, \qquad 
\mu^2=\frac{{\cal C}}{r^2} + \mu_0^2 \,,
\end{equation}
where the gluon PDF is found by a solution of the DGLAP evolution equation. The BGBK model accounts for the gluon splitting 
function $P_{gg}(z)$ only. The starting gluon PDF at the initial scale $\mu^2=\mu_0^2$ is parameterised as follows
\begin{eqnarray}
xg(x,\mu_0^2) = A_g x^{-\lambda_g} (1-x)^{5.6}\,, \quad \frac{\partial xg(x,\mu^2)}{\partial \ln \mu^2 } = 
\frac{\alpha_s(\mu^2)}{2\pi} \int_x^1 dz\, P_{gg}(z) \frac{x}{z} g\Big(\frac{x}{z}, \mu^2\Big)\,.
\label{dglap}
\end{eqnarray}
The parameters for the model were found by fitting the HERA data and read,
\begin{eqnarray}
A_g = 1.2\,, \quad \lambda_g = 0.28\,, \quad \mu_0^2 = 0.52\, \mathrm{GeV}^2\,, \quad 
{\cal C} = 0.26\,, \quad \sigma_0 = 23\, \mathrm{mb}\,.
\end{eqnarray}
\begin{figure*}[t]
\begin{minipage}{0.495\textwidth}
 \centerline{\includegraphics[width=1.12\textwidth]{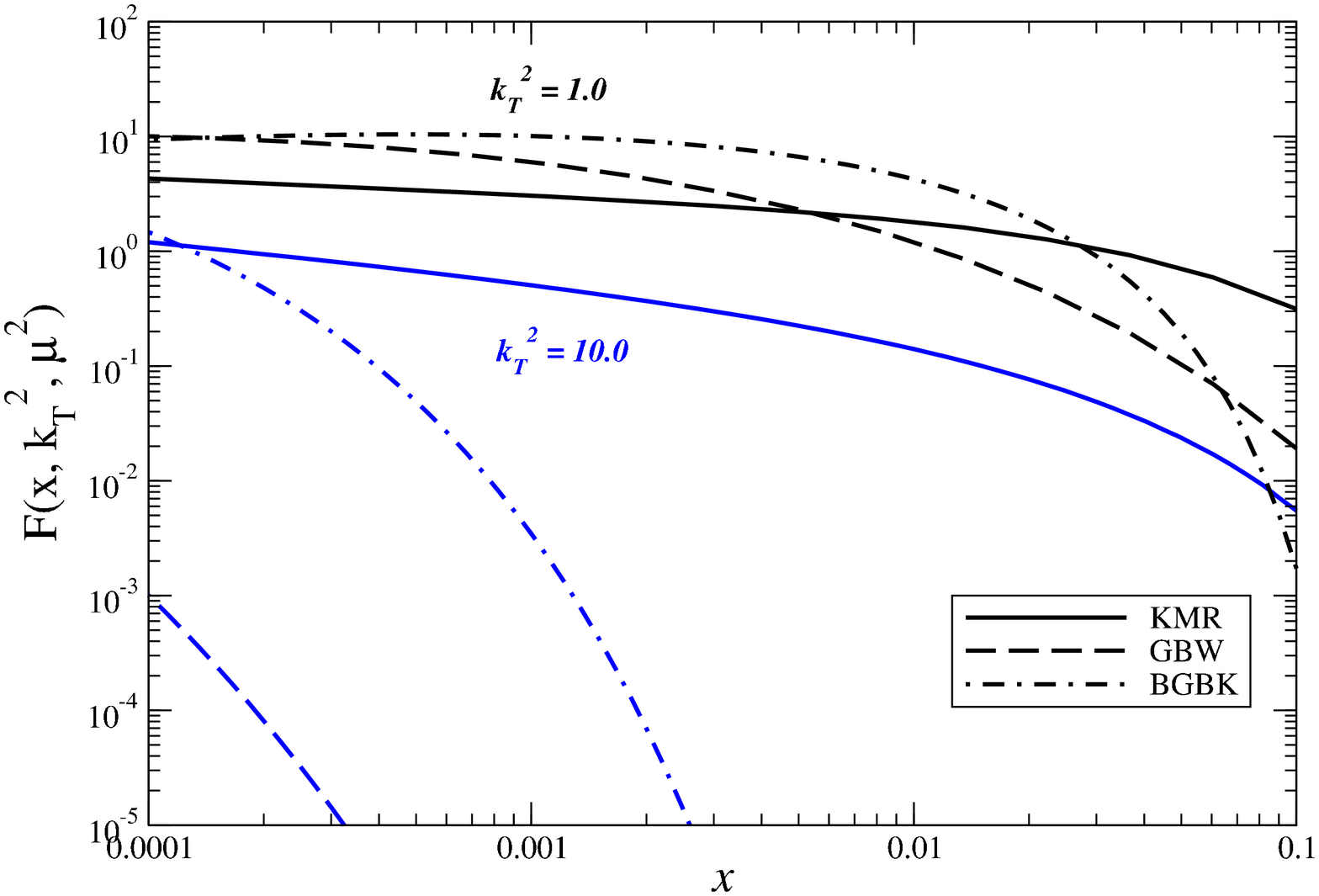}}
\end{minipage}
\begin{minipage}{0.495\textwidth}
 \centerline{\includegraphics[width=1.12\textwidth]{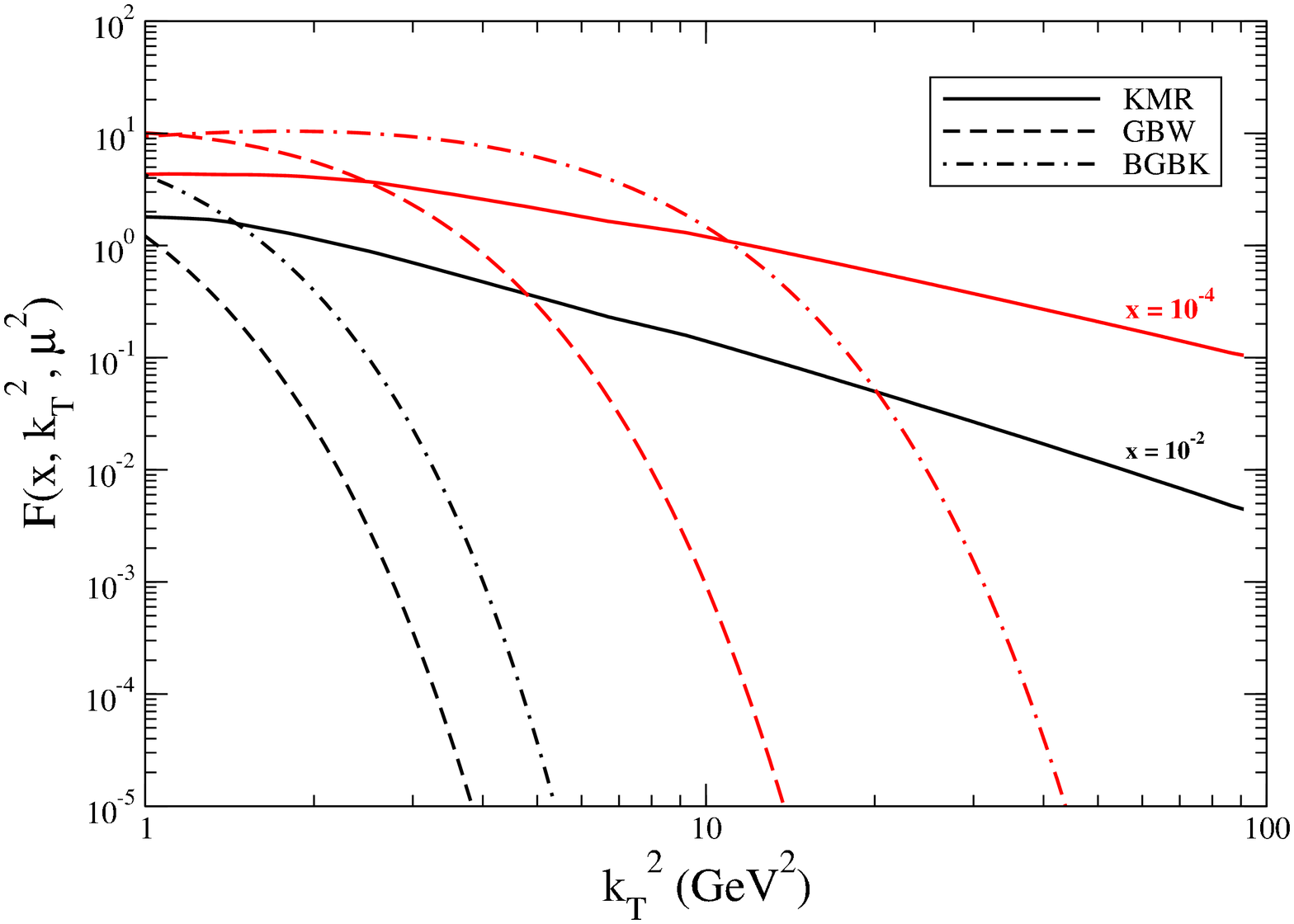}}
\end{minipage}
   \caption{ \small 
(Color online) The unintegrated KMR (solid lines), GBW (dashed lines) and BGBK (dash-dotted lines) gluon distributions in the target proton
as functions of the longitudinal momentum fraction $x$ at fixed $k_T^2=1,\,10$ GeV$^2$ values (left panel) and 
the transverse momentum squared $k_T^2$ at fixed $x=10^{-2},\,10^{-4}$ values (right panel).
}
\label{fig:unintegrated}
\end{figure*}

In Fig.~\ref{fig:unintegrated} for comparison we show the dipole TMD ${\cal K}_{\rm dip}(x,k_T^2)/\alpha_s(k_T^2)$ (\ref{dipole-UGDF}) 
and the conventional Kimber-Martin-Ryskin (KMR) UGDF model \cite{KMR}. The dipole TMD is based upon the GBW and BGBK 
parameterisations given by Eqs.~(\ref{gbw}) and (\ref{bgbk}), respectively, while the KMR model is constructed from the conventional quark 
and gluon densities and accounts for the coherent effects in gluon emissions corresponding to the main part of the collinear higher-order QCD 
corrections. Note that the KMR model is based on the standard DGLAP evolution not accounting for non-linear QCD effects. As expected these distributions 
depicted in Fig.~\ref{fig:unintegrated} exhibit very different $x$ and $k_T$ dependence. In particular, the GBW and BGBK distributions are exponentially 
suppressed at large values of $k_T$ and are enhanced at small transverse momenta while the KMR UGDF model has a power-like behavior. The suppression 
at large transverse momenta in the GBW and BGBK models is directly associated with the exponential saturated shape of the dipole cross section. 
Since the differences between the three models are so large, it is instructive to see how they imply for observables in comparison 
to the experimental data on $p_T$ spectra of heavy-flavored jets and mesons produced in high-energy $pp$ collisions.
\begin{figure*}[!h]
\begin{minipage}{0.495\textwidth}
 \centerline{\includegraphics[width=1.12\textwidth]{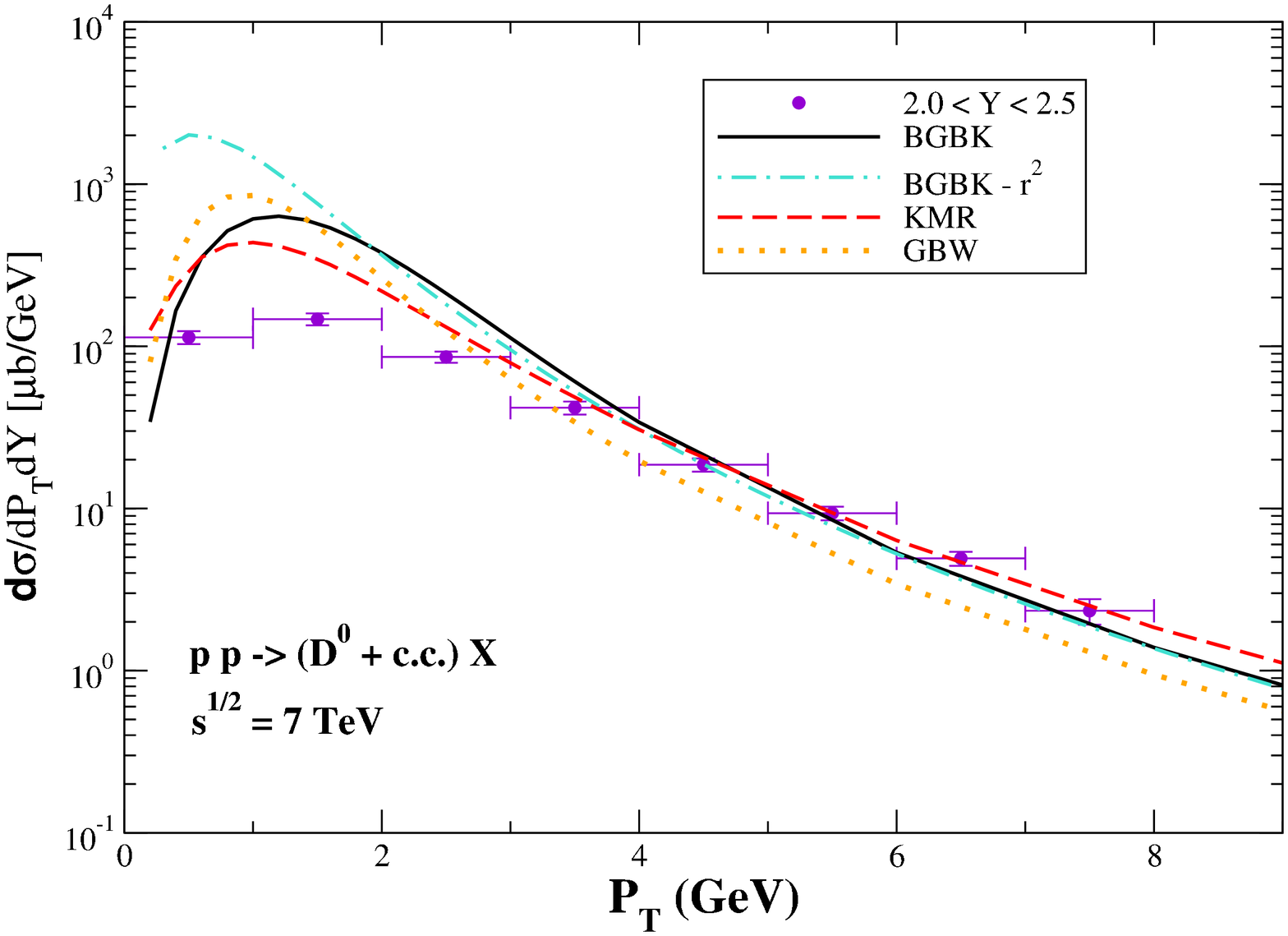}}
\end{minipage}
\begin{minipage}{0.495\textwidth}
 \centerline{\includegraphics[width=1.12\textwidth]{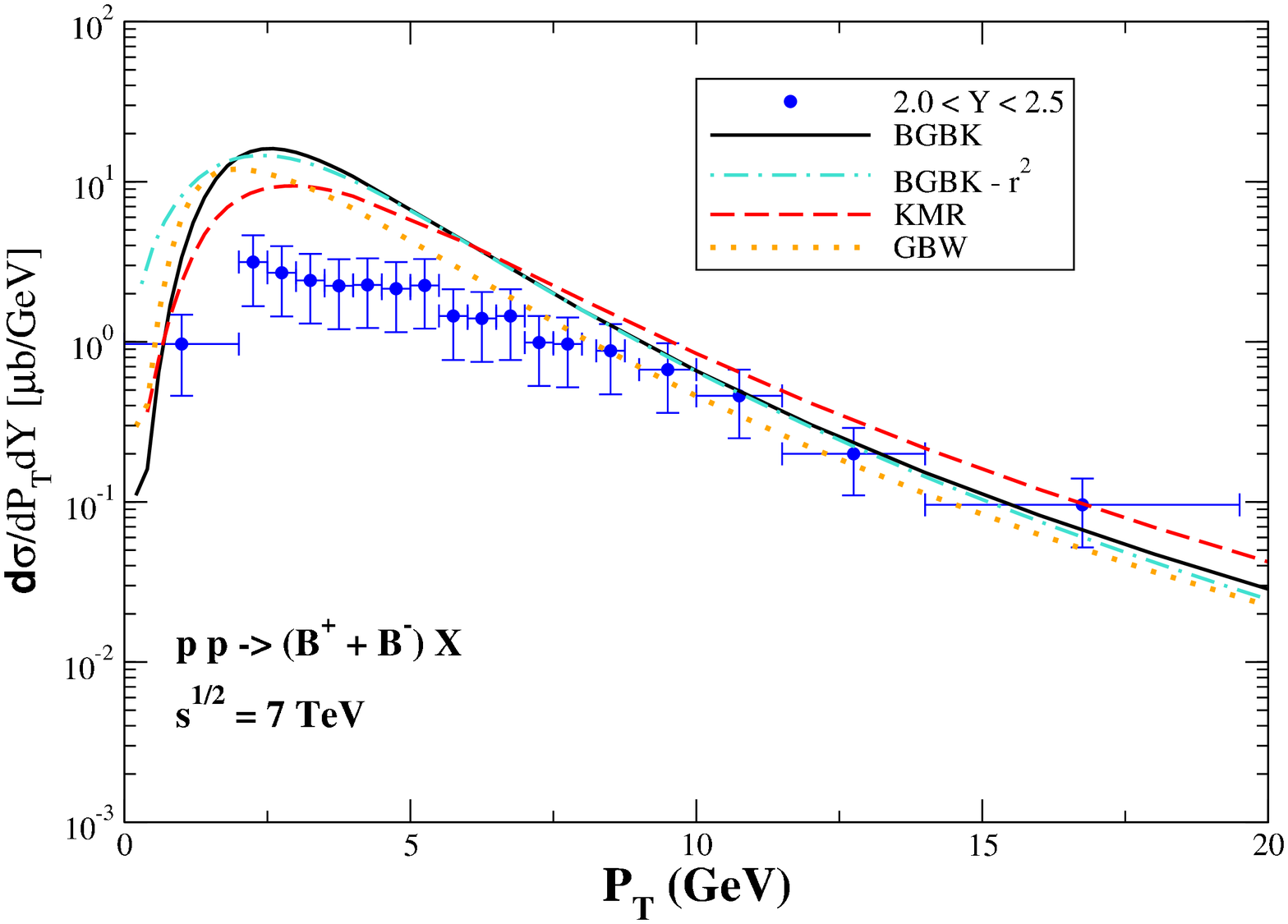}}
\end{minipage}
\begin{minipage}{0.495\textwidth}
 \centerline{\includegraphics[width=1.12\textwidth]{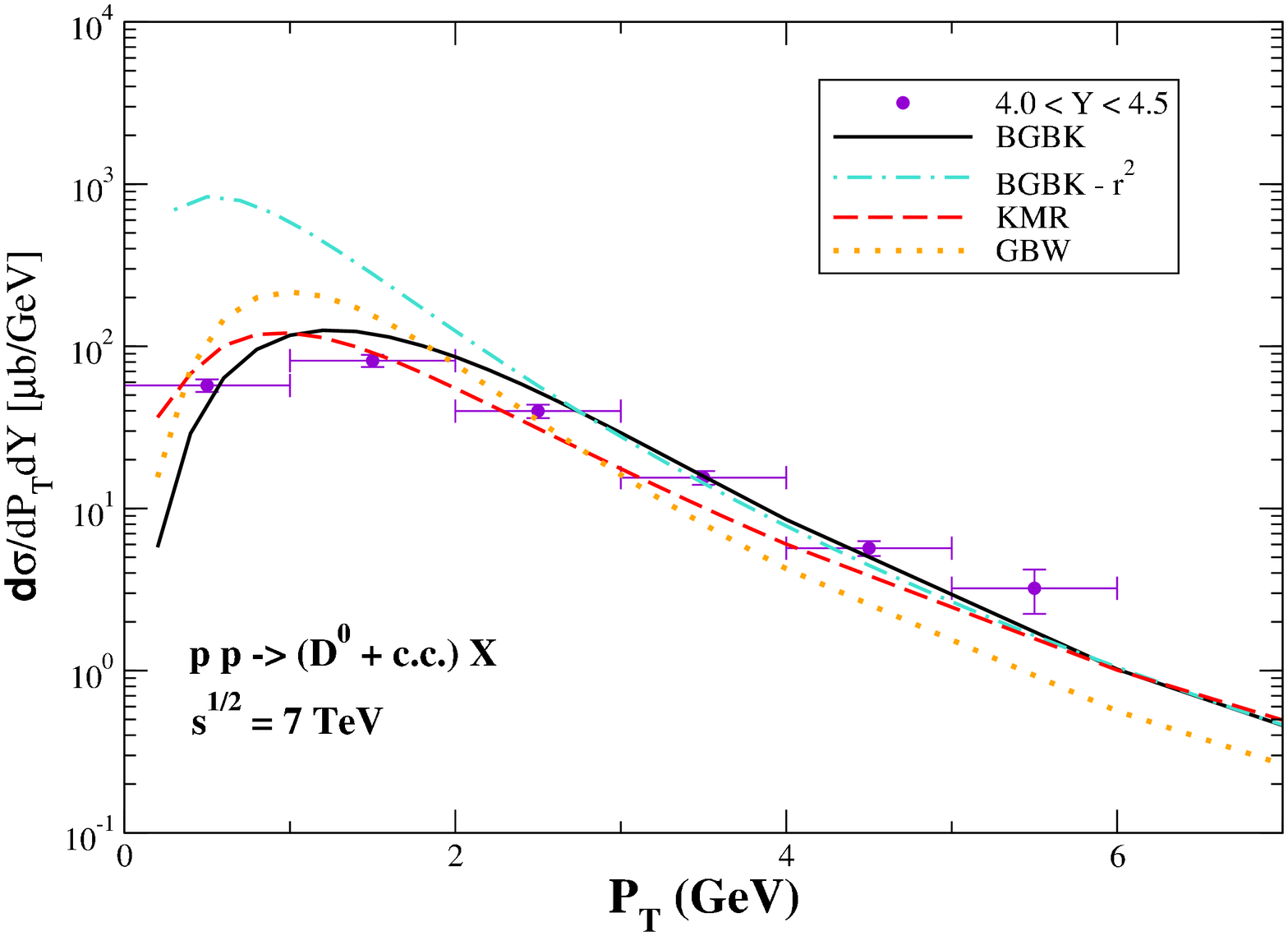}}
\end{minipage}
\begin{minipage}{0.495\textwidth}
 \centerline{\includegraphics[width=1.12\textwidth]{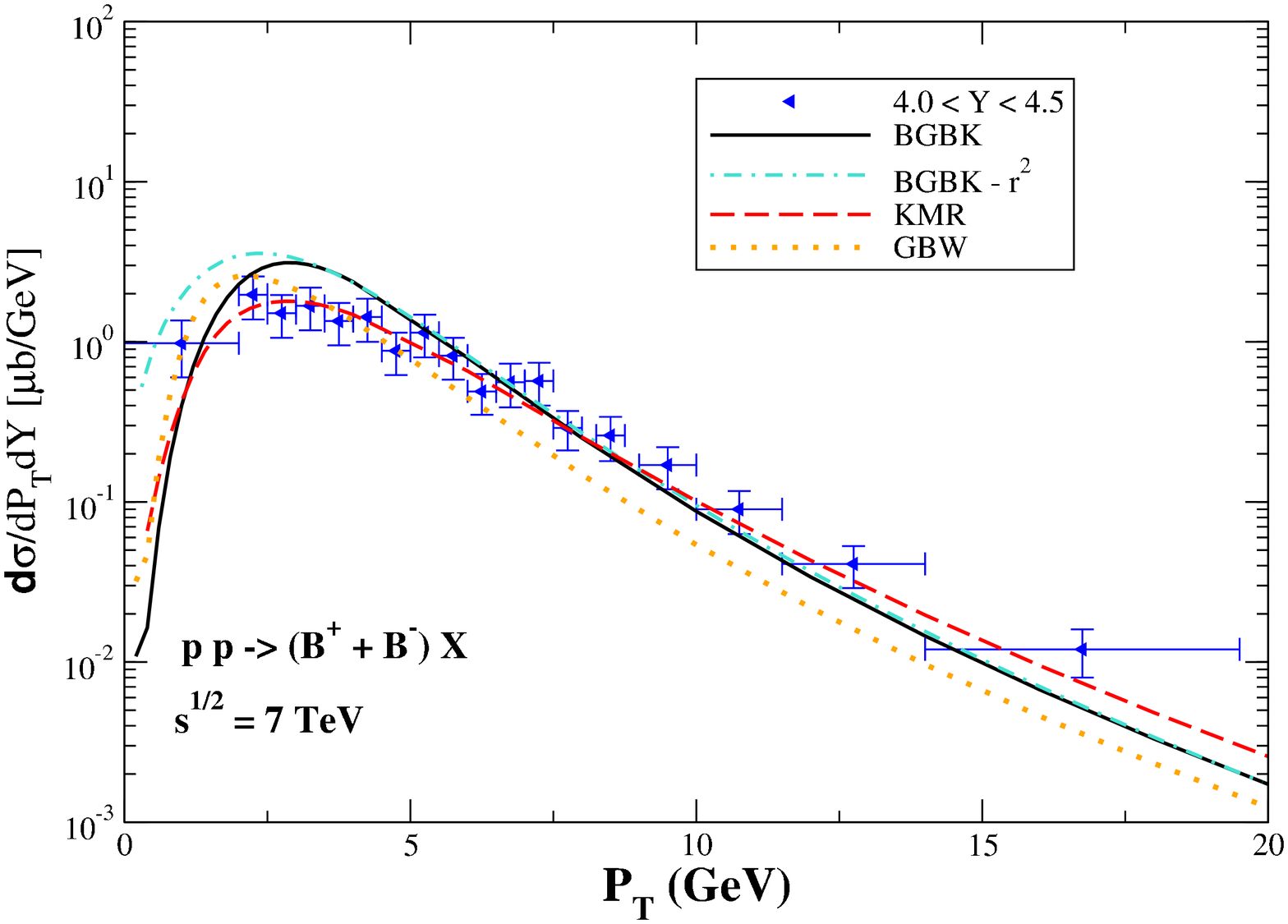}}
\end{minipage}
   \caption{ \small 
   (Color onine) The dipole model results for the differential $D^0$-meson (left panels) and $B^\pm$-meson (right panels) production cross sections in two distinct rapidity 
   bins, $2.0<Y<2.5$ (upper panels) and $4.0<Y<4.5$ (lower panels) as a function of meson transverse momentum $P_T$ versus the LHCb data at $\sqrt{s}=7$ TeV
   \cite{Dmesons_lhcb,Bmesons_lhcb}. Here, the results for the BGBK \cite{bgbk} (in saturated and in quadratic forms) and GBW \cite{GBW} dipole parameterisations 
   are compared to the result obtained with KMR UGDF \cite{KMR}. The collinear projectile gluon PDF in the CT10 model \cite{ct10} is adopted here. 
   }
 \label{fig:ALL-r2}
\end{figure*}

\section{Numerical results}
\label{Sec:results}

We are now in the position to calculate and consequently to discuss the numerical results for the differential cross section of open heavy flavor production  
obtained in the framework of dipole approach using Eqs.~(\ref{factkt}), (\ref{dipole-f}), (\ref{spectra:ms}) and (\ref{mes-CS}).

In Fig.~\ref{fig:ALL-r2} we show the differential $P_T$ distributions of $D^0$ (left panels) and $B^\pm$ mesons (right panels) produced 
in $pp$ collisions at c.m. energy $\sqrt{s}=7$ TeV versus data from the LHCb Collaboration \cite{Dmesons_lhcb,Bmesons_lhcb}. Such a comparison 
is shown for two well separated rapidity bins, $2.0<Y<2.5$ (upper panels) and $4.0<Y<4.5$ (lower panels). Performing the calculations, the GBW \cite{GBW} 
and BGBK \cite{bgbk} parameterisations for the dipole cross section (Eqs.~(\ref{gbw}) and (\ref{bgbk}), respectively) have been employed. The corresponding 
results are compared with those obtained using the KMR UGDF model assuming Eq.~(\ref{dip-vs-kt}).
\begin{figure*}[tbh]
\begin{minipage}{0.495\textwidth}
 \centerline{\includegraphics[width=1.12\textwidth]{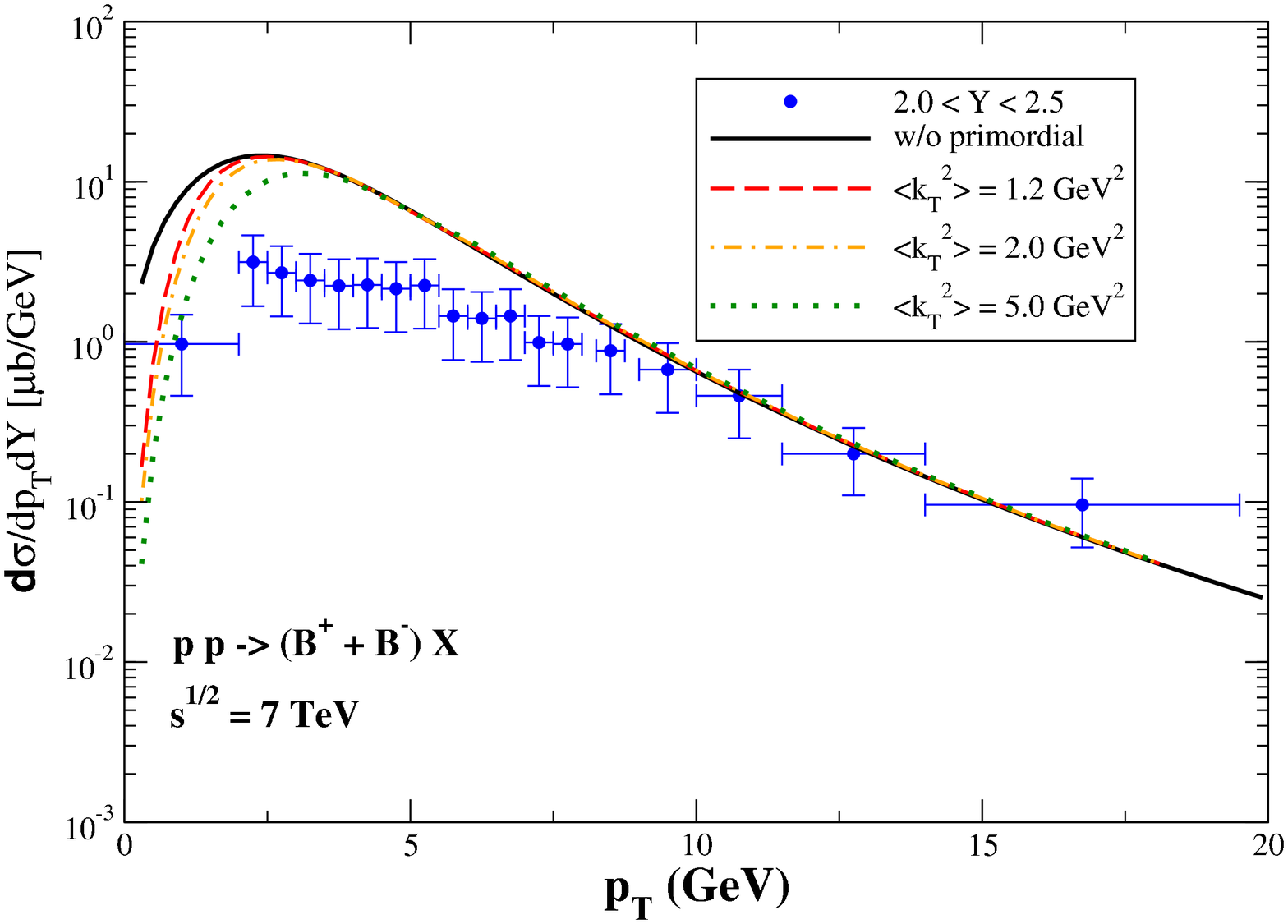}}
\end{minipage}
\begin{minipage}{0.495\textwidth}
 \centerline{\includegraphics[width=1.12\textwidth]{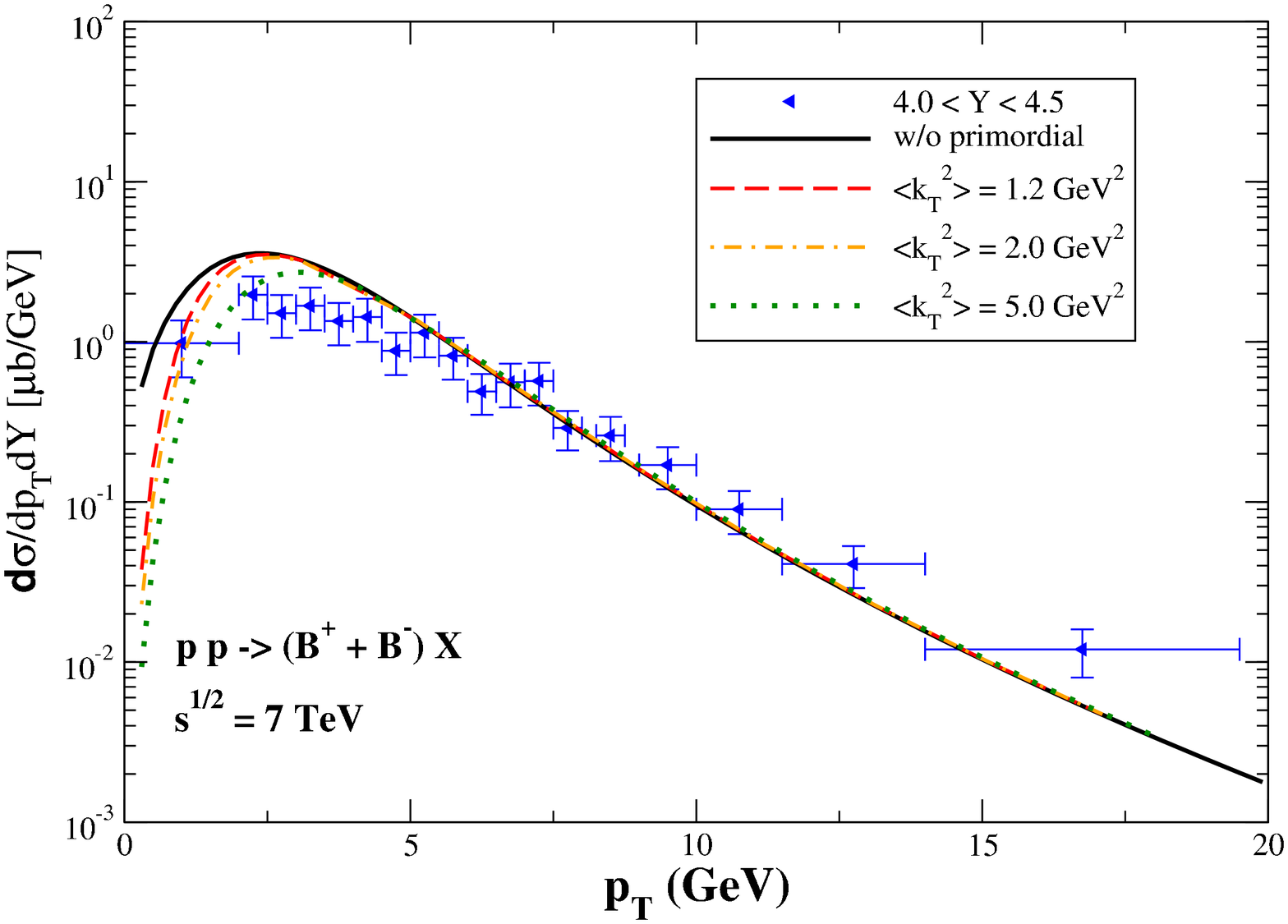}}
\end{minipage}
   \caption{ \small 
   (Color onine) The dipole model results for the differential $B^\pm$-meson production cross section in two distinct rapidity 
   intervals, $2.0<Y<2.5$ (left panel) and $4.0<Y<4.5$ (right panel), as a function of meson transverse momentum $P_T$ versus the LHCb data 
   at $\sqrt{s}=7$ TeV \cite{Bmesons_lhcb}. Here, the quadratic form of the BGBK model has been used for the target gluon while 
   the projectile gluon has been set to be collinear (solid line) or having a primordial transverse momentum treated via 
   the Gaussian smearing UGDF (\ref{Gauss}) with different values of the averaged $\langle k_T^2 \rangle = 1.0,\,2.0$ 
   and $5.0$ GeV$^2$ depicted by dashed, dash-dotted and dotted lines, respectively.
   }
 \label{fig:B-primG}
\end{figure*}

A good description of data is apparent at large $P_T\gtrsim 2m_Q$ using both the KMR and BGBK model, while the GBW model 
somewhat underestimates the data. However, the data at lower $P_T < 2m_Q$ are fairly well described only for the most 
forward rapidity insterval $4.0<Y<4.5$ especially for GBW and KMR models. For the most central rapidity bin $2.0<Y<2.5$ there is
a significant descrepancy with the data for $P_T\sim m_Q$ similar to all three dipole parameterisations.
Such low-$P_T$ behavior implies a rising significance of the primordial transverse momentum evolution of the projectile gluon density at central rapidities
which was not taken into consideration in Fig.~\ref{fig:ALL-r2}. The saturated shape of the dipole cross section (or the corresponding dipole TMD) plays a more pronounced 
role for lighter $D^0$-meson observables indicating a significant deviation of the approximate results using the quadractic form of $\sigma_{q\bar q}$ (\ref{CT}) 
from the saturated ansatz (\ref{saturation}). Note that a drammatic difference in the gluon $k_T$ shapes between the KMR and GBW UGDF models indicated 
in Fig.~\ref{fig:unintegrated} causes rather small differences in the $P_T$ spectra of the produced mesons.

The lack of agreement between our results and the experimental data at low meson $P_T$ values can be related to a primordial transverse momentum distribution 
of the projectile gluon in the incoming proton. The latter can be accounted by an additional convolution with the projectile gluon $k_T$-distribution as was done in 
Eq.~(\ref{factkt}). Indeed, in the framework of QCD parton model it was known since a long time ago that the experimental data on heavy quark \cite{Mangano:1991jk}, 
Drell-Yan \cite{Hom:1976zn,Kaplan:1977kr} and direct photon \cite{Apanasevich:1997hm,Apanasevich:1998ki} production at NLO can only be described if 
one incorporates an average primordial transverse momentum $\langle k_T^2\rangle \simeq 1$ GeV$^2$. Such a large value of $\langle k_T^2\rangle$ may 
indicate at a perturbative origin of the primordial momentum in the parton model. This situation makes it difficult to separate non-perturbative intrinsic and 
perturbatively-generated transverse momenta which is an open question in the QCD parton model.

In the framework of dipole approach, both perturbative and non-perturbative contributions to the intrinsic primordial parton momenta, except for a finite-size effect 
of the projectile hadron, are incorporated into the dipole cross section fitted to the DIS data. Thus, one should expect that the primordial transverse momentum
of the projectile gluon in the dipole picture could have essentially a non-perturbative nature \cite{Kopeliovich:2007yva}. So, $\langle k_T^2\rangle$ should be 
considerably less than found in the QCD parton model. In this case, 
an intrinsic primordial momentum distribution can be accounted using Eq.~(\ref{factkt}) and assuming a model for the unintegrated gluon distribution 
of the incident gluon.
\begin{figure*}[!h]
\begin{minipage}{0.495\textwidth}
 \centerline{\includegraphics[width=1.12\textwidth]{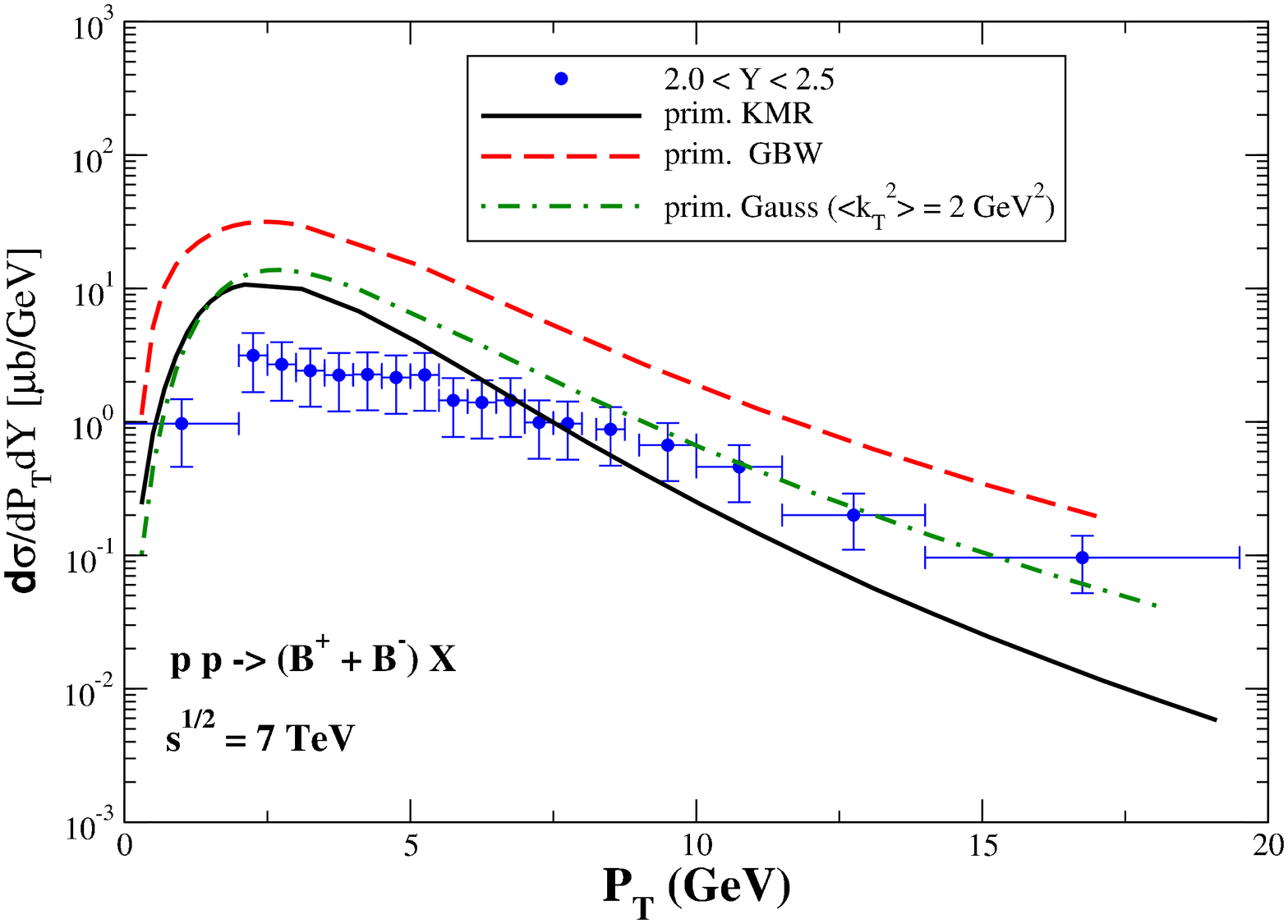}}
\end{minipage}
\begin{minipage}{0.495\textwidth}
 \centerline{\includegraphics[width=1.12\textwidth]{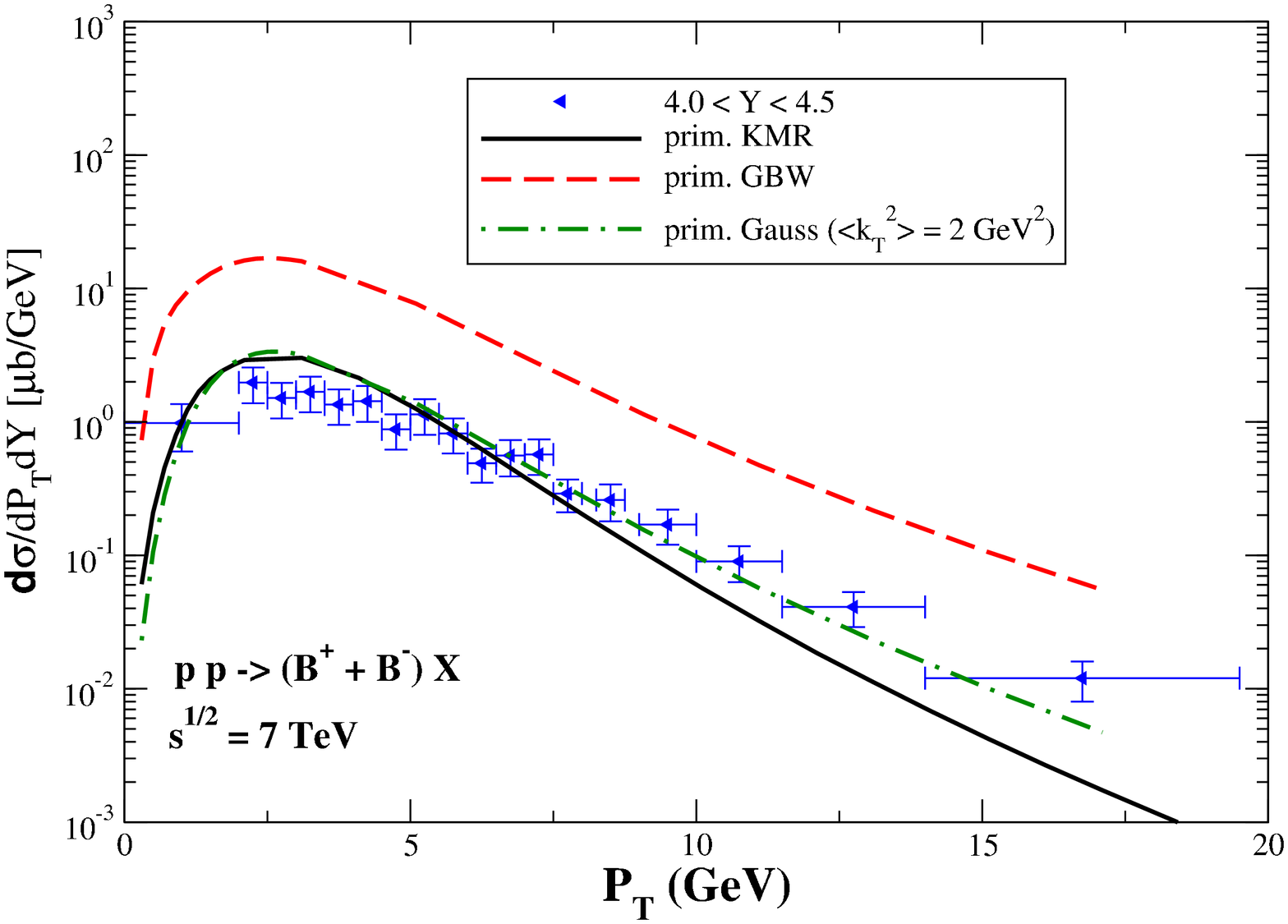}}
\end{minipage}
   \caption{ \small 
   (Color onine) A comparison of dipole model predictions for differential $B$-meson production cross section using different models of the primordial UGDF with 
   the corresponding LHCb data \cite{Bmesons_lhcb} in two distinct rapidity intervals. Here, the quadratic form of the BGBK model has been used for the target gluon. 
   The projectile gluon UGDF has been chose to be KMR (solid line), GBW (dashed line) and Gaussian smearing (dash-dotted line) with $\langle k_T^2 \rangle = 2.0$ GeV$^2$.}
\label{fig:prim}
\end{figure*}

In order to check the expectation that in the dipole picture the intrinsic primordial momentum is small, in Fig.~\ref{fig:B-primG} we show the $P_T$ spectra 
of produced $B^\pm$ mesons in the dipole framework compared to the LHCb data \cite{Bmesons_lhcb} in two distinct rapidity intervals 
$2.0<Y<2.5$ (left panel) and $4.0<Y<4.5$ (right panel). Here, we consider a Gaussian smearing model for the UGDF, where the intrinsic transverse momentum
of the distribution can be factorized and is smeared by a normalized Gaussian distribution given by
\begin{eqnarray} 
\label{Gauss}
{\cal G}_N(k_T)=\frac{1}{\pi \langle k_T^2\rangle}\,e^{-k_T^2/\langle k_T^2\rangle} \,.
\end{eqnarray}
The results are shown for different values of the averaged $\langle k_T^2 \rangle = 1.0,\,2.0$ and $5.0$ GeV$^2$ depicted by dashed, dash-dotted and dotted lines, 
respectively. We notice that for $B^{\pm}$ mesons the impact of intrinsic $k_T$ on meson $P_T$ spectra is small within the interval of $\langle k_T^2 \rangle$ used
in our calculations. This observation indicates the perturbative origin of the intrinsic $k_T$-dependence of the projectile gluon UGDF. 

In Fig.~\ref{fig:prim} we have compared the predictions for three different primordial UGDF models for the projectile gluon $k_T$ distribution: KMR (solid line), GBW (dashed line) 
and Gaussian smearing according to Eq.~(\ref{Gauss}) (dash-dotted line) with $\langle k_T^2 \rangle = 2.0$ GeV$^2$. We see that none of the models is able to improve 
the data description at low $P_T<2m_Q$ and for $2.0<Y<2.5$. We expect much larger $\langle k_T^2 \rangle \sim m_Q^2$ in order to obtain a better description 
in the small $P_T$ region. At the same time, such large values of $\langle k_T^2 \rangle$ imply that the use of the non-perturbative $k_T$ distribution (\ref{Gauss}) 
is not applicable anymore. Fig.~\ref{fig:prim} also shows that the KMR and Gaussian smearing models predict a rather similar magnitude of the cross section at low $P_T$.
However, the primordial KMR UGDF significantly underestimates the data at large $P_T$ due to relative enhancement of large gluon $k_T$ values compares to other 
dipole parameterisations. The primordial gluon $k_T$ evolution predicted by $k_T$ factorisation in such models as KMR is not in correspondence with typical dipole 
parameterisations. Due this reason, in particular, the primordial GBW UGDF overestimates the data by an order of magnitude. In addition, the GBW model is not applicable 
at large $x_1\gtrsim 0.01$. Such inconsistency with low-$P_T$ data arises the question about the properties of $k_T$ evolution of the primordial gluon density in 
the dipole picture. To summarise, none of the popular phenomenological models for the primordial UGDF can reproduce the data in the range of low $P_T$ and $Y$. 
The dipole model provides so an important tool for constraining the primordial UGDFs using all available data.
\begin{figure*}[!h]
\begin{minipage}{0.495\textwidth}
 \centerline{\includegraphics[width=1.12\textwidth]{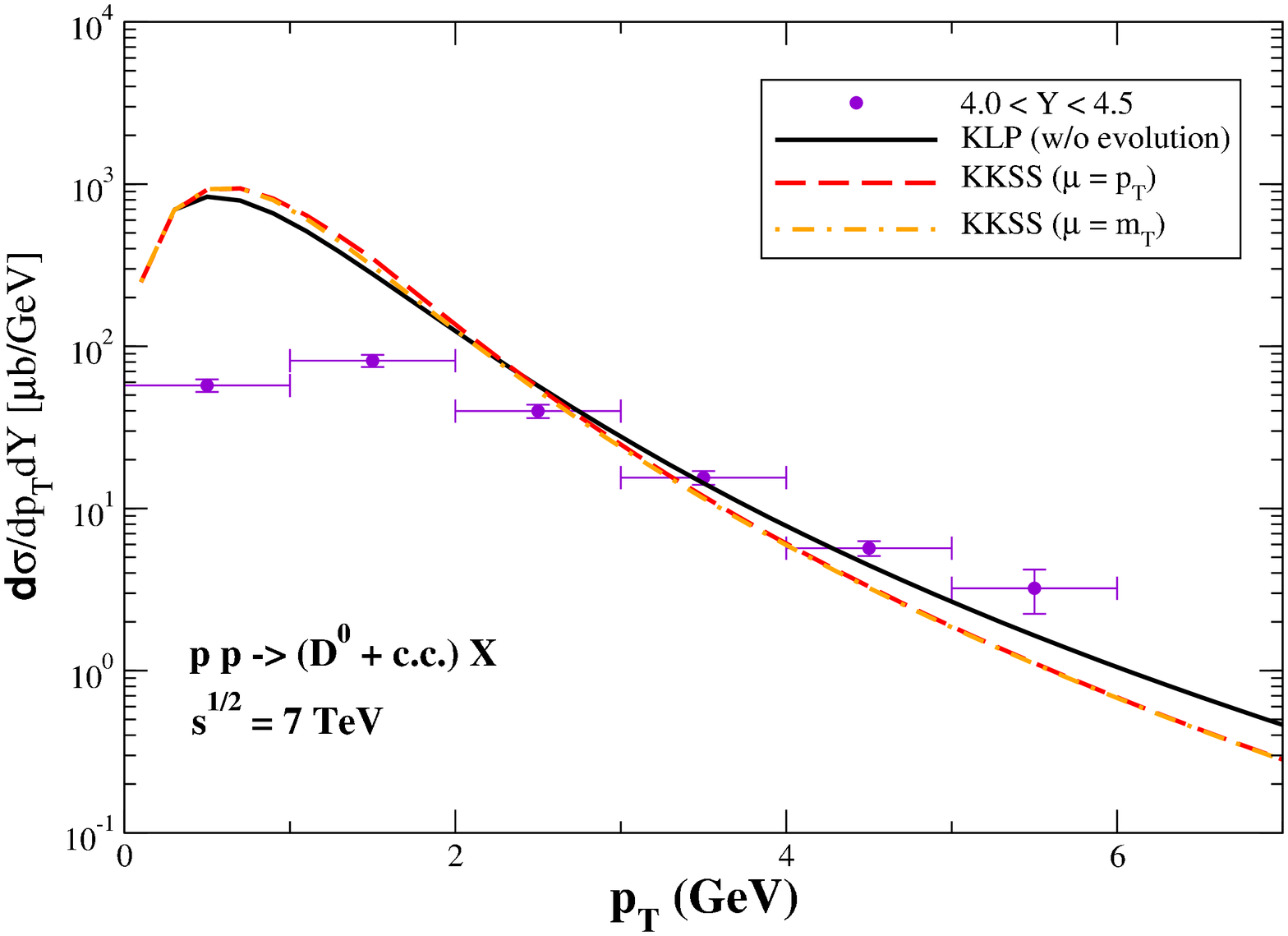}}
\end{minipage}
\begin{minipage}{0.495\textwidth}
 \centerline{\includegraphics[width=1.12\textwidth]{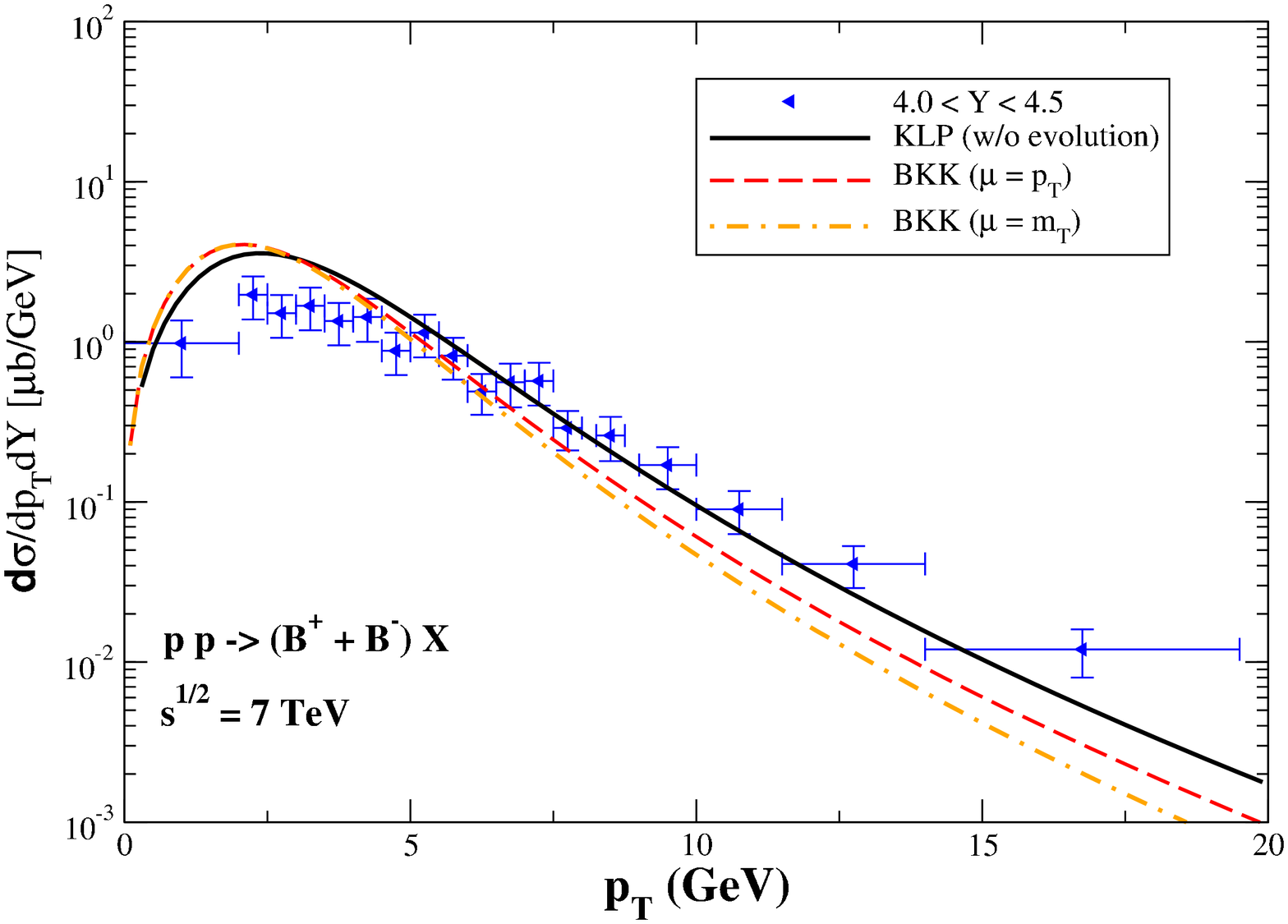}}
\end{minipage}
   \caption{ \small 
   (Color onine) The effect of QCD evolution in the fragmentation function on the differential $D^0$-meson (left panel) 
   and $B^\pm$-meson (right panel) production cross sections. Here, the quadratic form of the BGBK model has been 
   used for the target gluon. Data are taken from the LHCb Collaboration \cite{Dmesons_lhcb,Bmesons_lhcb}.}
 \label{fig:FF}
\end{figure*}

Fig.~\ref{fig:FF} clearly demonstrates the importance of the onset of QCD evolution in the fragmentation functions. Here, we compare our predictions for
the differential cross sections for $D^0$ (left panel) and $B^\pm$ (right panel) meson production with the corresponding LHCb data \cite{Dmesons_lhcb,Bmesons_lhcb}. 
At low $P_T\lesssim 5$ GeV, the standard Kartvelishvili-Likhoded-Petrov (KLP) parameterisation of fragmentation functions \cite{KLP} provides sufficiently precise results. 
As expected, the importance of the DGLAP evolution increases with $P_T$. The BKK model \cite{BKK} gives rise to a suppression of $B^\pm$ mesons at large $P_T$ compared 
to the KLP result.
\begin{figure*}[!h]
\begin{minipage}{0.495\textwidth}
 \centerline{\includegraphics[width=1.12\textwidth]{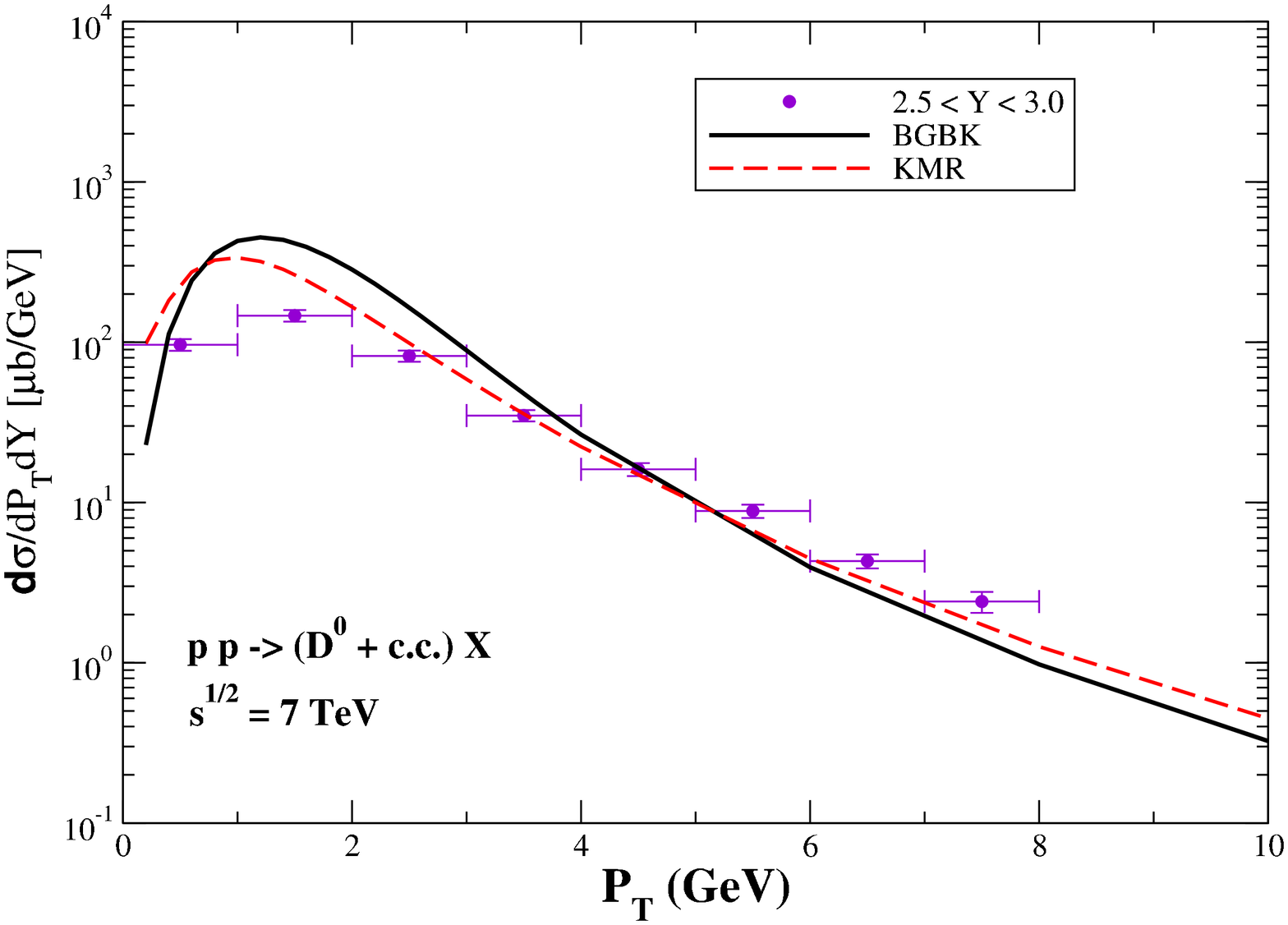}}
\end{minipage}
\begin{minipage}{0.495\textwidth}
 \centerline{\includegraphics[width=1.12\textwidth]{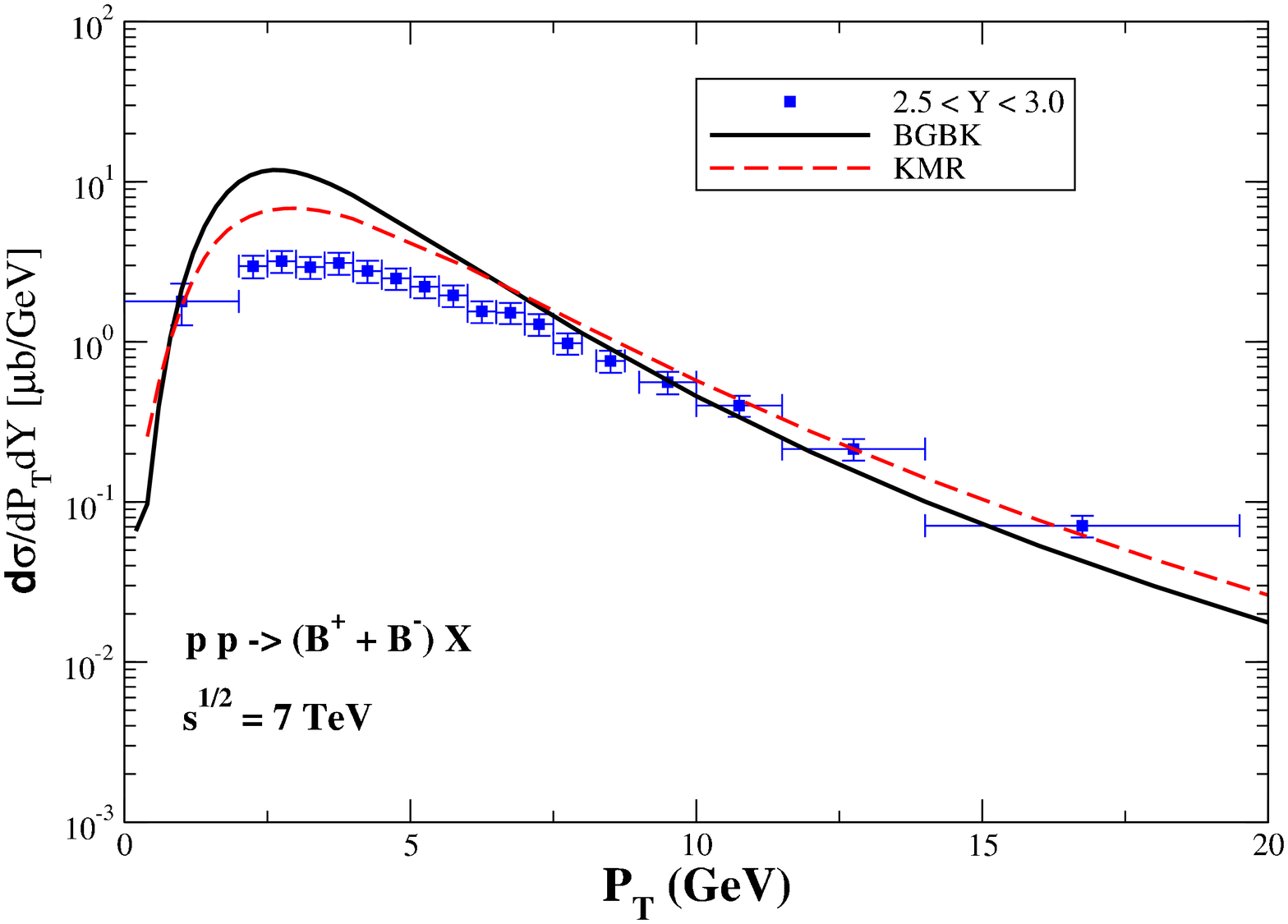}}
\end{minipage}
\begin{minipage}{0.495\textwidth}
 \centerline{\includegraphics[width=1.12\textwidth]{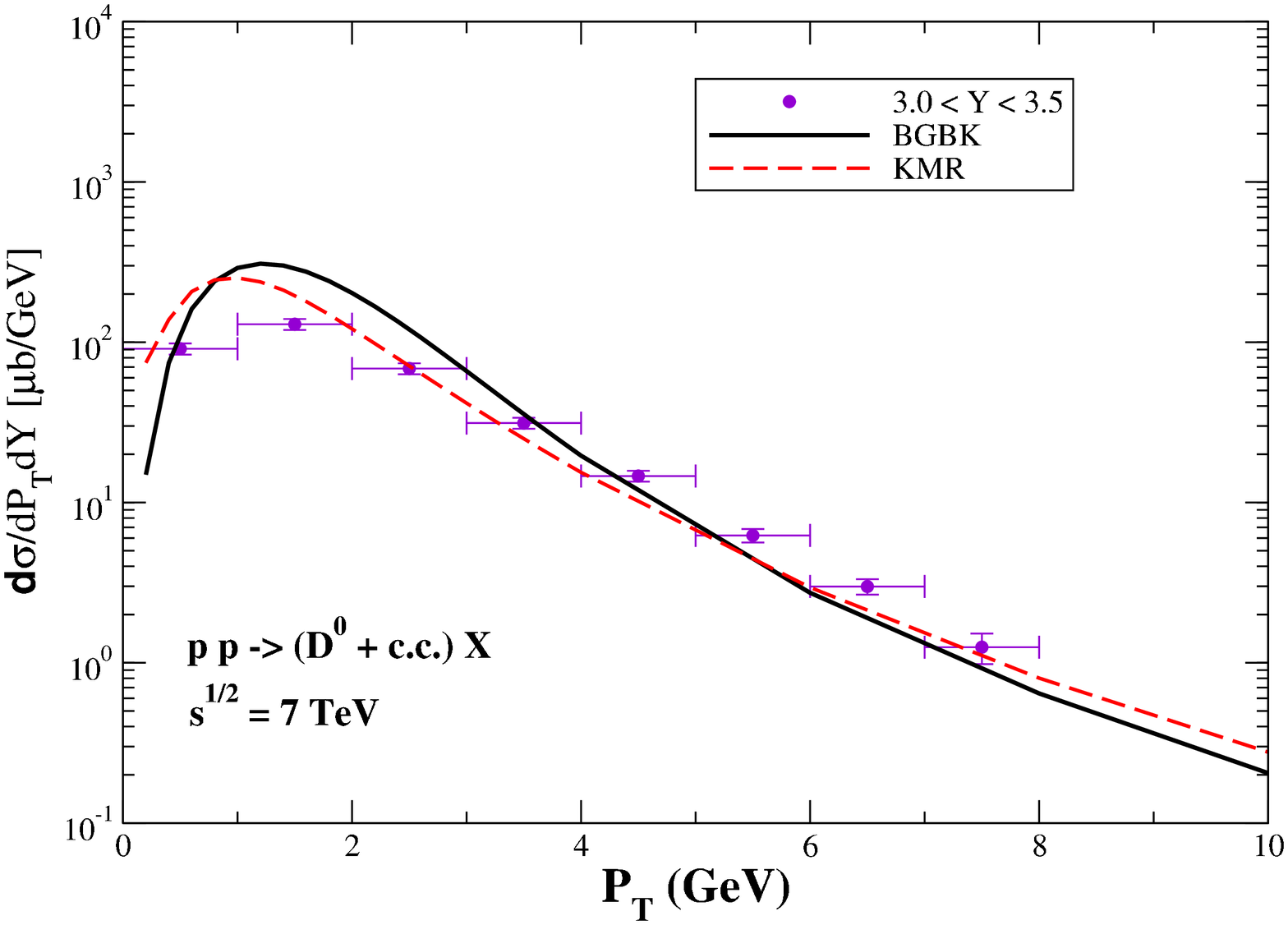}}
\end{minipage}
\begin{minipage}{0.495\textwidth}
 \centerline{\includegraphics[width=1.12\textwidth]{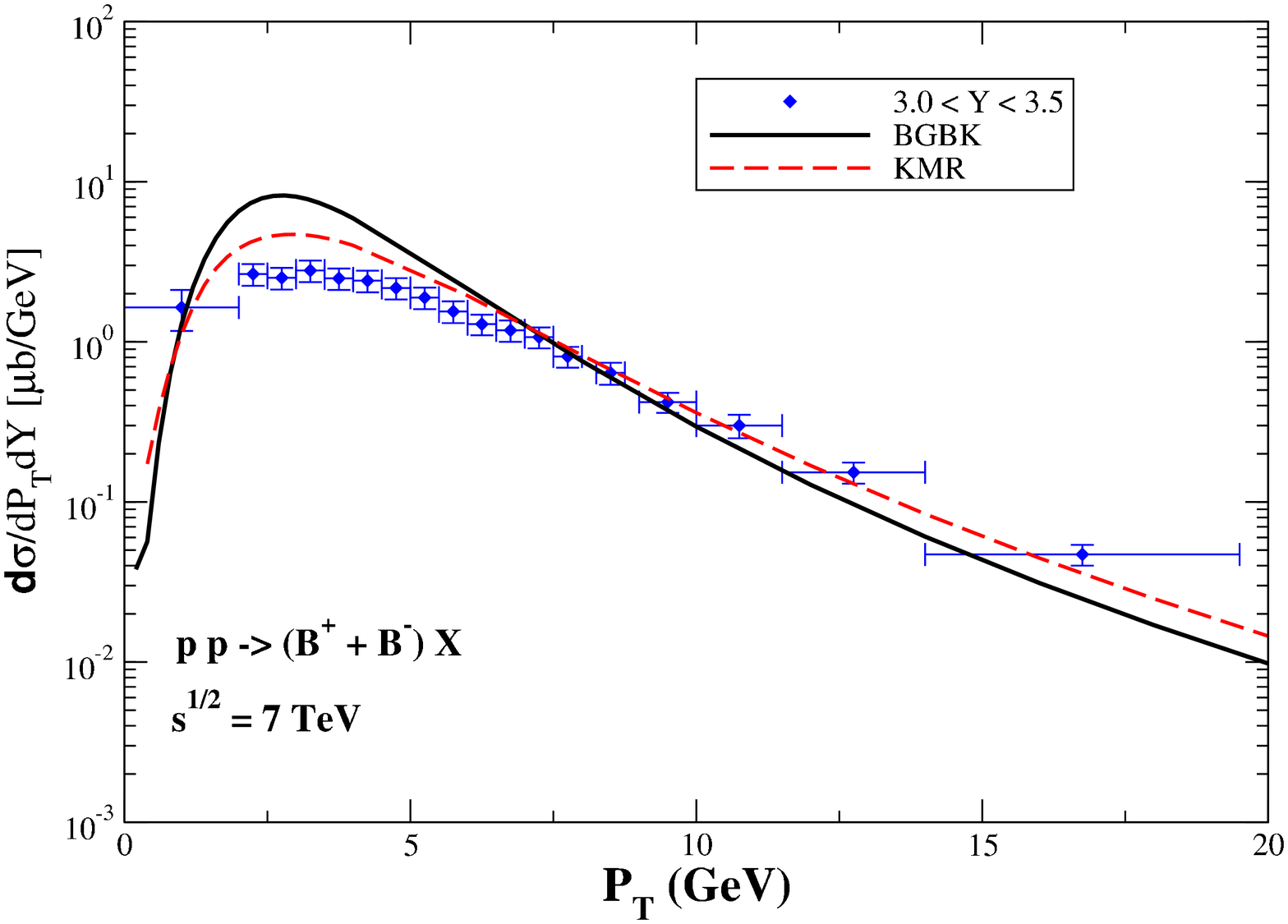}}
\end{minipage}
\begin{minipage}{0.495\textwidth}
 \centerline{\includegraphics[width=1.12\textwidth]{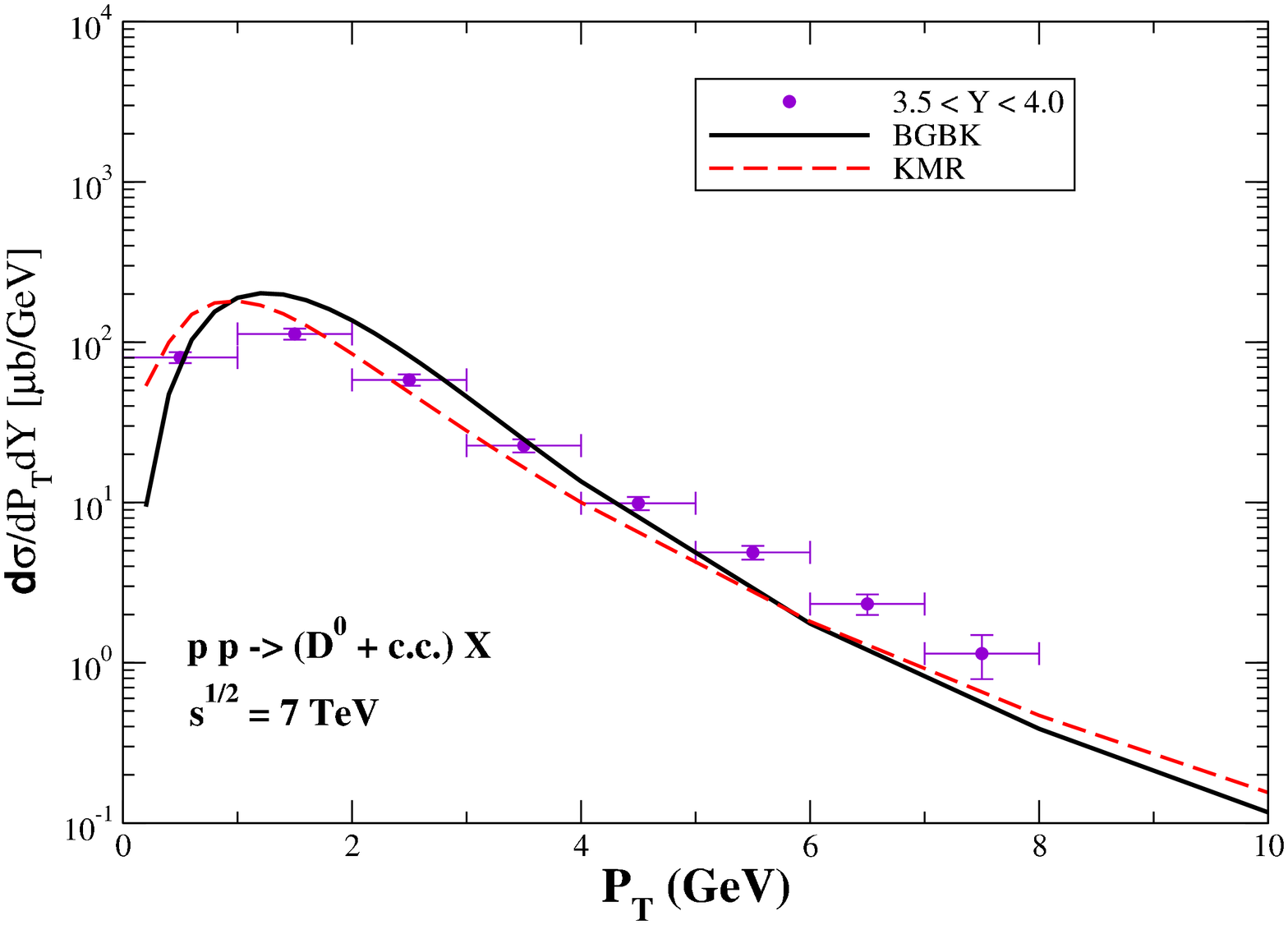}}
\end{minipage}
\begin{minipage}{0.495\textwidth}
 \centerline{\includegraphics[width=1.12\textwidth]{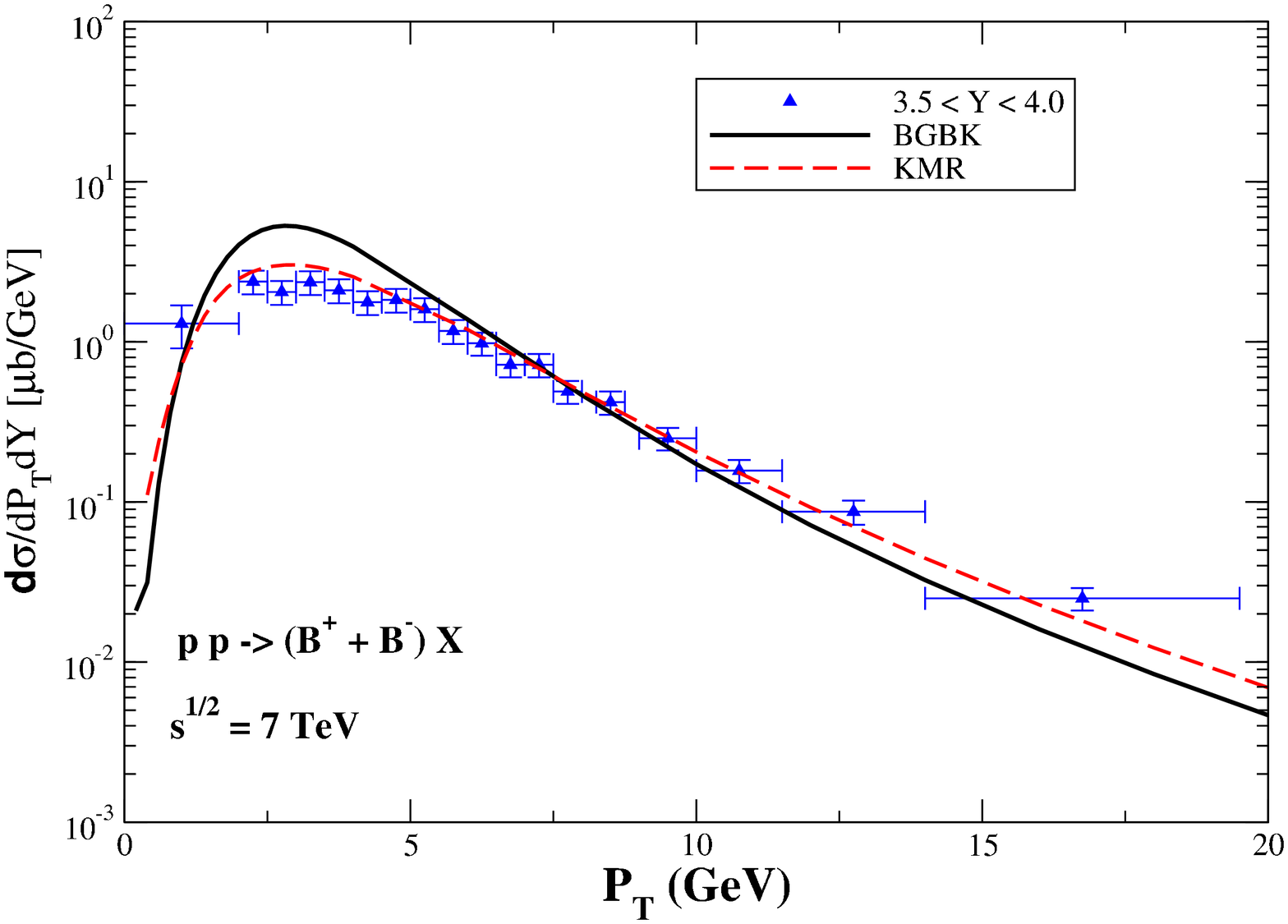}}
\end{minipage}
   \caption{ \small 
   (Color online) The dipole model predictions with KMR UGDF and the BGBK model (\ref{bgbk}) for transverse momentum distributions 
   of $D^0$ (left panels) and $B^\pm$ (right panels) mesons in various rapidity intervals at $\sqrt{s}=7$ TeV versus data taken from 
   the LHCb Collaboration \cite{Dmesons_lhcb,Bmesons_lhcb}.
   }
 \label{fig:LHCb-Y}
\end{figure*}

In Fig.~\ref{fig:LHCb-Y} we demonstrate how good the LHCb data \cite{Dmesons_lhcb,Bmesons_lhcb} the $D^0$ (left panels) and $B^\pm$ (right panels) meson 
production cross sections are described using the BGBK dipole model and KMR UGDF. Generally, the more forward rapidities are considered, the better is description 
of the data using both models. Note that in spite of absence of saturation effects in the KMR UGDF model, Fig.~\ref{fig:LHCb-Y} shows that this model provides suprisingly good 
description of the data even at low $P_T$ and large $Y$ values where the strong onset of saturation effects is expected.
\begin{figure*}[!h]
\begin{minipage}{0.495\textwidth}
 \centerline{\includegraphics[width=1.12\textwidth]{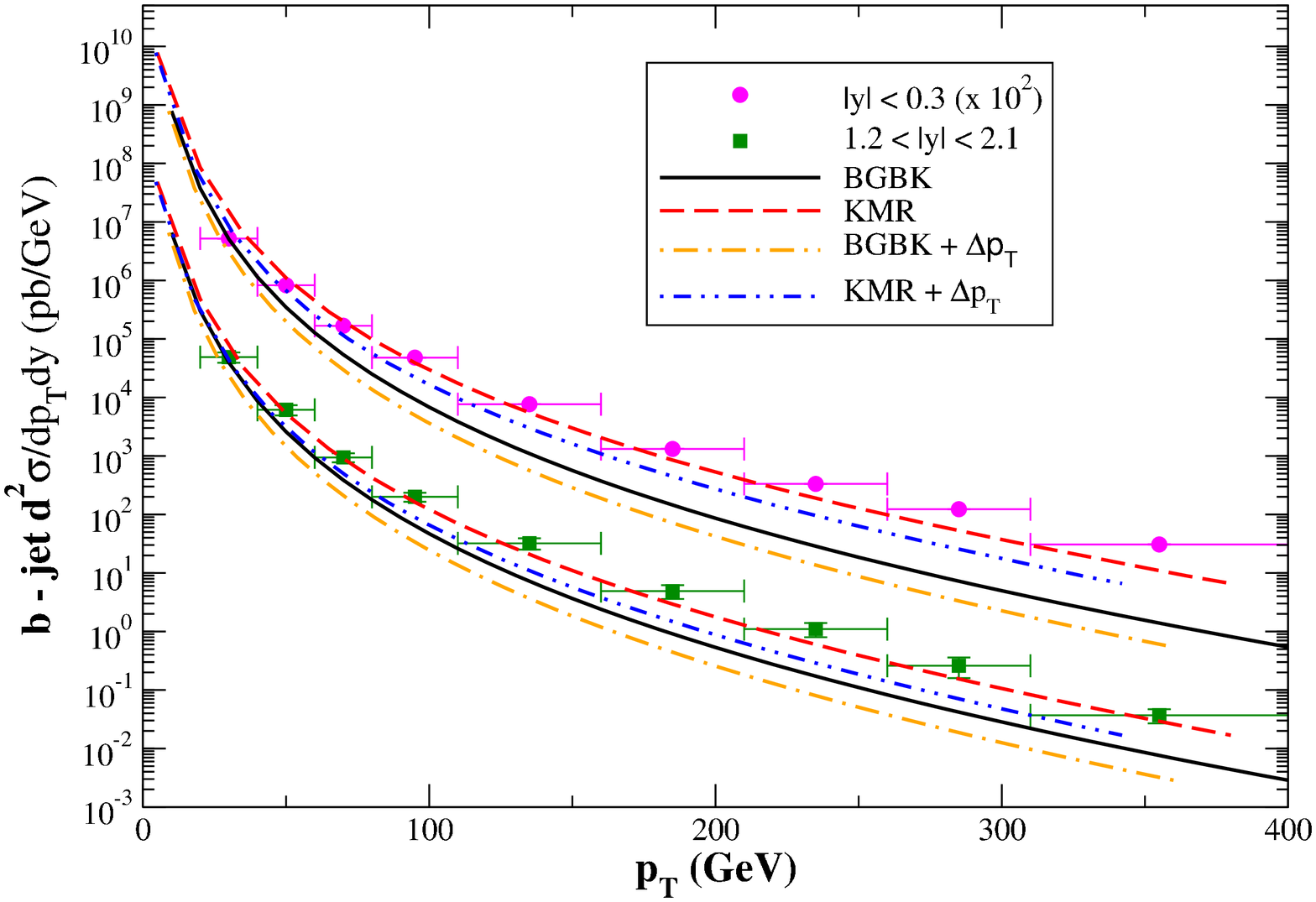}}
\end{minipage}
\begin{minipage}{0.495\textwidth}
 \centerline{\includegraphics[width=1.12\textwidth]{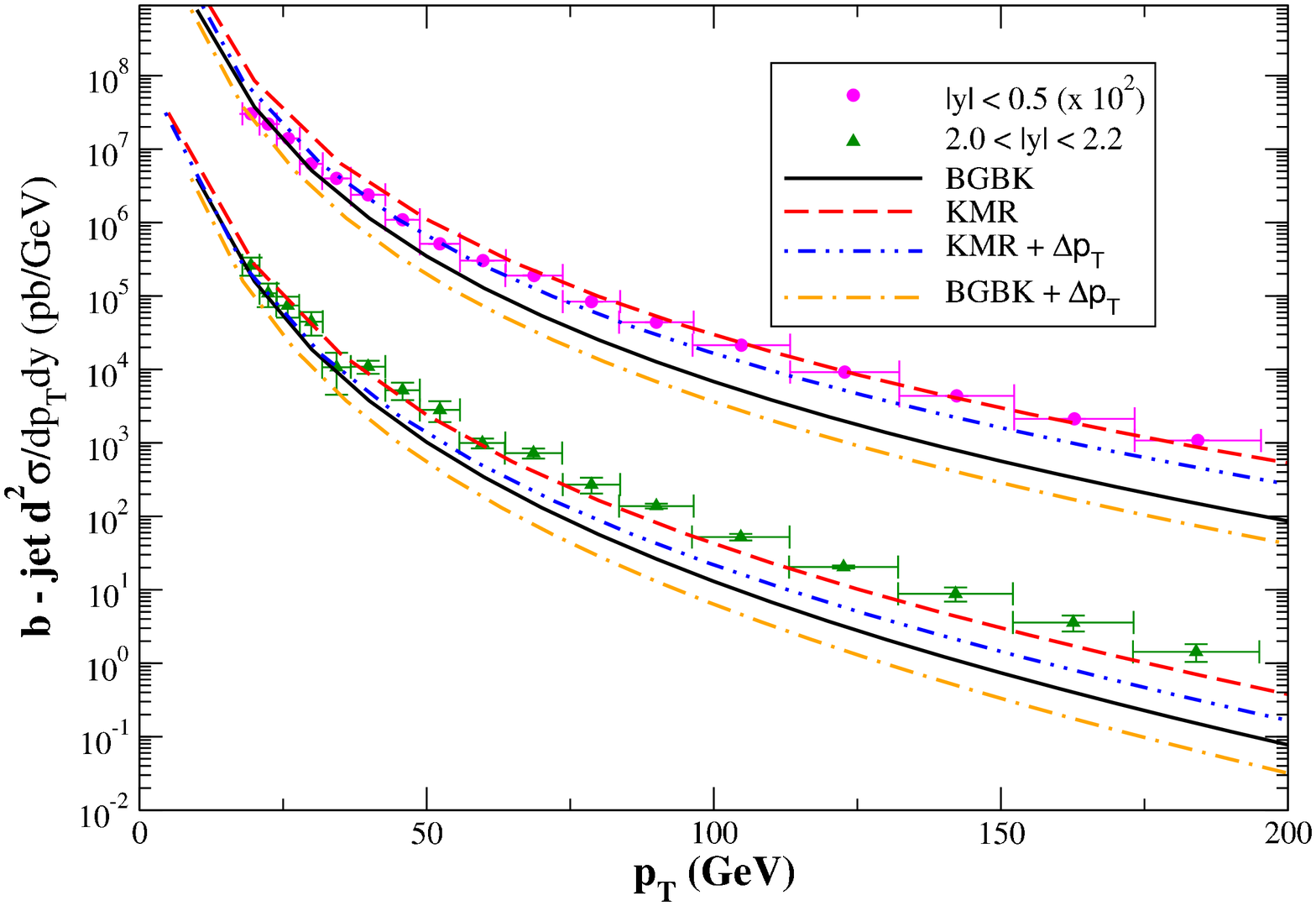}}
\end{minipage}
   \caption{ \small 
   (Color online) The dipole model predictions with KMR UGDF and the $r^2$-approximated BGBK model (\ref{bgbk}) for transverse 
   momentum distributions of $b$-jets at $\sqrt{s}$=7 TeV integrated over two distinct rapidity intervals. The data are taken from the ATLAS Collaboration 
   for $|y|<0.3$, $1.2<|y|<2.1$ \cite{atlas_bjets} (left panel) and from the CMS Collaboration for $|y| < 0.5$, $2.0<|y|<2.2$~\cite{cms_bjets} (right panel).
   }
 \label{fig:bjet}
\end{figure*}

Finally, Fig.~\ref{fig:bjet} shows a comparison between the dipole model results for the differential $b$-jet production cross section as a function of the jet transverse 
momentum $p_T$ with the corresponding ATLAS \cite{atlas_bjets} (left panel) and CMS \cite{cms_bjets} (right panel) data. Here, we present these results 
in comparison with experimental data for two distinct rapidity intervals only as is depicted in Fig.~\ref{fig:bjet}. We note here that the jet energy leakage effect described 
in Section~\ref{Sec:E-loss} leads to a reduction of the jet production cross section. In ATLAS measurements \cite{atlas_bjets}, the jet radius parameter is $R=0.4$, 
while at CMS \cite{cms_bjets} $R=0.5$ is adopted independent on jet rapidity. Using Fig.~\ref{fig:jet-cone} with such values of $R$, we found the corresponding relative 
shifts in jet transverse momentum $\Delta p_T / p_T$, caused by the gluon radiation outside of the jet cone, and implemented them in Fig.~\ref{fig:bjet}.
Accounting for the resulting reduction of the jet cross section, one observes that the KMR UGDF model (dash-dot-dotted lines) describes the ATLAS data reasonably 
well in the whole interval of $p_T$ while it somewhat underestimates the CMS data. The quadratic form of BGBK model (dash-dotted lines) significantly underestimates 
the data, both from ATLAS and CMS, especially at large $p_T>50$ GeV. One therefore concludes that the LHC $b$-jet data at large $p_T$ represent an effective probe 
enabling us to test QCD evolution of the saturation scale.  

\section{Summary}
\label{Sec:summary}

In conclusion, we have analysed the most recent LHC data on the differential (in transverse momentum and rapidity) cross sections for open heavy flavor production
in the framework of color dipole model. 

We demonstrate that at large values of heavy meson transverse momenta $P_T\gtrsim 2m_Q$, $Q=c,b$ and/or forward rapidities $Y>3.5$, the dipole model predictions 
employing the GBW and BGBK dipole parameterisations, as well as the KMR UGDF in the target nucleon, are generally consistent with the available data, with a few exceptions. 
Namely, in the low $P_T$ region $P_T<2m_Q$ at central rapidities $Y<3.5$, the data are not well described indicating so a significant role of the primordial intrinsic transverse momentum dependence of the projectile gluon density. The use of conventional UGDF models for the primordial gluon density does not improve the data description but
demonstrates significant theoretical uncertainties in these kinematic regions. Despite of the fact that the KMR UGDF does not account for the saturation effects it describes 
the heavy flavored ($D^0$ and $B^\pm$) meson data surprisingly well at large rapidities $Y>3.5$ in both low and high $P_T$ domains. As for the $b$-jet differential 
distributions, with an account for the jet energy leakage effect due to gluon radiation outside the jet cone, the use of KMR UGDF in the target gluon density leads to 
a reasonably good description of the ATLAS data in the whole region of $p_T<400$ GeV while it somewhat underestimates the CMS data. The BGBK model noticeably 
underestimates the ATLAS and CMS data, especially at large $p_T>50$ GeV.

The dipole approach thus provides an efficient tool for analysis of the heavy flavor hadroproduction at the LHC. It directly accesses and could potentially be used to constrain 
such phenomena as saturation dynamics in $pp$ collisions, initial-state evolution in primordial $k_T$ and hadronisation of heavy quarks.
\vspace{0.5cm}

{\bf Acknowledgments} Stimulating discussions with M. Siddikov are acknowledged. V.P.G. is supported by CNPq, CAPES and FAPERGS, Brazil. B.K. and 
I.P. are supported by Fondecyt (Chile) grants No. 1140377 and 1170319, as well as by USM-TH-342 grant and by CONICYT grant PIA ACT1406 (Chile). 
J.N. is partially supported by the grant 13-20841S of the Czech Science Foundation (GA\v CR), by the Grant M\v SMT LG15001, by the Slovak Research 
and Development Agency APVV-0050-11 and by the Slovak Funding Agency, Grant 2/0020/14. R.P. is partially supported by the Swedish Research Council, 
contract number 621-2013-428 and by CONICYT grant PIA ACT1406 (Chile).



\begin{thebibliography}{99}

\bibitem{Frixione:1997ma} 
  S.~Frixione, M.~L.~Mangano, P.~Nason and G.~Ridolfi,
  Adv.\ Ser.\ Direct.\ High Energy Phys.\  {\bf 15}, 609 (1998).

\bibitem{Baines:2006uw} 
  J.~Baines {\it et al.},
  hep-ph/0601164.

\bibitem{Andronic:2015wma} 
  A.~Andronic {\it et al.},
  Eur.\ Phys.\ J.\ C {\bf 76}, no. 3, 107 (2016).

\bibitem{Collins:1988ig} 
  J.~C.~Collins, D.~E.~Soper and G.~F.~Sterman,
  Nucl.\ Phys.\ B {\bf 308}, 833 (1988).

\bibitem{Collins:1989gx} 
  J.~C.~Collins, D.~E.~Soper and G.~F.~Sterman,
  Adv.\ Ser.\ Direct.\ High Energy Phys.\  {\bf 5}, 1 (1989).

\bibitem{Nason:1987xz} 
  P.~Nason, S.~Dawson and R.~K.~Ellis,
  Nucl.\ Phys.\ B {\bf 303}, 607 (1988).

\bibitem{Altarelli:1988qr} 
  G.~Altarelli, M.~Diemoz, G.~Martinelli and P.~Nason,
  Nucl.\ Phys.\ B {\bf 308}, 724 (1988).

\bibitem{Beenakker:1988bq} 
  W.~Beenakker, H.~Kuijf, W.~L.~van Neerven and J.~Smith,
  Phys.\ Rev.\ D {\bf 40}, 54 (1989).

\bibitem{Nason:1989zy} 
  P.~Nason, S.~Dawson and R.~K.~Ellis,
  Nucl.\ Phys.\ B {\bf 327}, 49 (1989)
  Erratum: [Nucl.\ Phys.\ B {\bf 335}, 260 (1990)].

\bibitem{Beenakker:1990maa} 
  W.~Beenakker, W.~L.~van Neerven, R.~Meng, G.~A.~Schuler and J.~Smith,
  Nucl.\ Phys.\ B {\bf 351}, 507 (1991).

\bibitem{levin91}
  E.~M.~Levin, M.~G.~Ryskin, Yu.~M.~Shabelski and A.~G.~Shuvaev,
  Sov.\ J.\ Nucl.\ Phys.\  {\bf 53}, 657 (1991)
  [Yad.\ Fiz.\  {\bf 53}, 1059 (1991)].

\bibitem{levin92}
  E.~M.~Levin, M.~G.~Ryskin, Yu.~M.~Shabelski and A.~G.~Shuvaev,
  Sov.\ J.\ Nucl.\ Phys.\  {\bf 54}, 867 (1991)
  [Yad.\ Fiz.\  {\bf 54}, 1420 (1991)].

\bibitem{Gribov:1972ri} 
  V.~N.~Gribov and L.~N.~Lipatov,
  Sov.\ J.\ Nucl.\ Phys.\  {\bf 15}, 438 (1972)
  [Yad.\ Fiz.\  {\bf 15}, 781 (1972)].

\bibitem{Altarelli:1977zs} 
  G.~Altarelli and G.~Parisi,
  Nucl.\ Phys.\ B {\bf 126}, 298 (1977).

\bibitem{Dokshitzer:1977sg} 
  Y.~L.~Dokshitzer,
  Sov.\ Phys.\ JETP {\bf 46}, 641 (1977)
  [Zh.\ Eksp.\ Teor.\ Fiz.\  {\bf 73}, 1216 (1977)].

\bibitem{Collins:1986mp} 
  J.~C.~Collins and W.~K.~Tung,
  Nucl.\ Phys.\ B {\bf 278}, 934 (1986).

\bibitem{Olness:1987ep} 
  F.~I.~Olness and W.~K.~Tung,
  Nucl.\ Phys.\ B {\bf 308}, 813 (1988).

\bibitem{Aivazis:1993kh} 
  M.~A.~G.~Aivazis, F.~I.~Olness and W.~K.~Tung,
  Phys.\ Rev.\ D {\bf 50}, 3085 (1994).

\bibitem{Aivazis:1993pi} 
  M.~A.~G.~Aivazis, J.~C.~Collins, F.~I.~Olness and W.~K.~Tung,
  Phys.\ Rev.\ D {\bf 50}, 3102 (1994).

\bibitem{Buza:1995ie} 
  M.~Buza, Y.~Matiounine, J.~Smith, R.~Migneron and W.~L.~van Neerven,
  Nucl.\ Phys.\ B {\bf 472}, 611 (1996).

\bibitem{Buza:1996wv} 
  M.~Buza, Y.~Matiounine, J.~Smith and W.~L.~van Neerven,
  Eur.\ Phys.\ J.\ C {\bf 1}, 301 (1998).

\bibitem{Martin:1996eva} 
  A.~D.~Martin, R.~G.~Roberts, M.~G.~Ryskin and W.~J.~Stirling,
  Eur.\ Phys.\ J.\ C {\bf 2}, 287 (1998).

\bibitem{Thorne:1997ga} 
  R.~S.~Thorne and R.~G.~Roberts,
  Phys.\ Rev.\ D {\bf 57}, 6871 (1998).

\bibitem{Thorne:1997uu} 
  R.~S.~Thorne and R.~G.~Roberts,
  Phys.\ Lett.\ B {\bf 421}, 303 (1998).

\bibitem{Cacciari:1998it} 
  M.~Cacciari, M.~Greco and P.~Nason,
  JHEP {\bf 9805}, 007 (1998)

\bibitem{Collins:1998rz} 
  J.~C.~Collins,
  Phys.\ Rev.\ D {\bf 58}, 094002 (1998).

\bibitem{Kniehl:2004fy} 
  B.~A.~Kniehl, G.~Kramer, I.~Schienbein and H.~Spiesberger,
  Phys.\ Rev.\ D {\bf 71}, 014018 (2005).

\bibitem{Kniehl:2005mk} 
  B.~A.~Kniehl, G.~Kramer, I.~Schienbein and H.~Spiesberger,
  Eur.\ Phys.\ J.\ C {\bf 41}, 199 (2005).

\bibitem{gribov83}
  L.~V.~Gribov, E.~M.~Levin and M.~G.~Ryskin,
  Phys.\ Rept.\  {\bf 100}, 1 (1983).

\bibitem{marchesini88}
  G.~Marchesini and B.~R.~Webber,
  Nucl.\ Phys.\  B {\bf 310}, 461 (1988); 
  Nucl.\ Phys.\  B {\bf 386}, 215 (1992).

\bibitem{levin90}
  E.~M.~Levin and M.~G.~Ryskin,
  Phys.\ Rept.\  {\bf 189}, 267 (1990).

\bibitem{cata90}
  S.~Catani, M.~Ciafaloni and F.~Hautmann,
  Phys.\ Lett.\  B {\bf 242}, 97 (1990).

\bibitem{cata91}
  S.~Catani, M.~Ciafaloni and F.~Hautmann,
  Nucl.\ Phys.\  B {\bf 366}, 135 (1991).

\bibitem{collins91}
  J.~C.~Collins and R.~K.~Ellis,
  Nucl.\ Phys.\  B {\bf 360}, 3 (1991).

\bibitem{ryskin99}
  M.~G.~Ryskin, A.~G.~Shuvaev and Yu.~M.~Shabelski,
  Phys.\ Atom.\ Nucl.\  {\bf 64}, 120 (2001)
  [Yad.\ Fiz.\  {\bf 64}, 123 (2001)]
  [arXiv:hep-ph/9907507].

\bibitem{hagler}
  P.~Hagler, R.~Kirschner, A.~Schafer, L.~Szymanowski and O.~Teryaev, 
  Phys.\ Rev.\  D {\bf 62}, 071502 (2000).

\bibitem{ryskin01}
  M.~G.~Ryskin, A.~G.~Shuvaev and Yu.~M.~Shabelski, 
  Phys.\ Atom.\ Nucl.\  {\bf 64}, 1995 (2001)
  [Yad.\ Fiz.\  {\bf 64}, 2080 (2001)]
  [arXiv:hep-ph/0007238].

\bibitem{shab2004}
  Yu.~M.~Shabelski and A.~G.~Shuvaev,
  Phys.\ Atom.\ Nucl.\  {\bf 69}, 314 (2006).

\bibitem{saleev} 
  V.~Saleev and A.~Shipilova,
  Phys.\ Rev.\ D {\bf 86}, 034032 (2012).

\bibitem{Chachamis:2015ona} 
  G.~Chachamis, M.~De\'ak, M.~Hentschinski, G.~Rodrigo and A.~Sabio Vera,
  JHEP {\bf 1509}, 123 (2015).

\bibitem{Collins:2007nk} 
  J.~Collins and J.~W.~Qiu,
  Phys.\ Rev.\ D {\bf 75}, 114014 (2007).

\bibitem{Rogers:2010dm} 
  T.~C.~Rogers and P.~J.~Mulders,
  Phys.\ Rev.\ D {\bf 81}, 094006 (2010).

\bibitem{DelDuca:2011ae} 
  V.~Del Duca, C.~Duhr, E.~Gardi, L.~Magnea and C.~D.~White,
  JHEP {\bf 1112}, 021 (2011)

\bibitem{Kopeliovich:2005ym} 
  B.~Z.~Kopeliovich, J.~Nemchik, I.~K.~Potashnikova, M.~B.~Johnson and I.~Schmidt,
  Phys.\ Rev.\ C {\bf 72}, 054606 (2005).

\bibitem{Nikolaev:1994de} 
  N.~N.~Nikolaev, G.~Piller and B.~G.~Zakharov,
  J.\ Exp.\ Theor.\ Phys.\  {\bf 81}, 851 (1995)
  [Zh.\ Eksp.\ Teor.\ Fiz.\  {\bf 108}, 1554 (1995)].

\bibitem{Nikolaev:1995ty} 
  N.~N.~Nikolaev, G.~Piller and B.~G.~Zakharov,
  Z.\ Phys.\ A {\bf 354}, 99 (1996).

\bibitem{Kopeliovich:2001ee} 
  B.~Kopeliovich, A.~Tarasov and J.~H\"ufner,
  Nucl.\ Phys.\ A {\bf 696}, 669 (2001).

\bibitem{Kopeliovich:2002yv} 
  B.~Z.~Kopeliovich and A.~V.~Tarasov,
  Nucl.\ Phys.\ A {\bf 710}, 180 (2002).

\bibitem{Kopeliovich:1981pz} 
  B.~Z.~Kopeliovich, L.~I.~Lapidus and A.~B.~Zamolodchikov,
  JETP Lett.\  {\bf 33}, 595 (1981)
  [Pisma Zh.\ Eksp.\ Teor.\ Fiz.\  {\bf 33}, 612 (1981)].

\bibitem{Nikolaev:1993th} 
  N.~N.~Nikolaev and B.~G.~Zakharov,
  Z.\ Phys.\ C {\bf 64}, 631 (1994).

\bibitem{Nikolaev:1994kk} 
  N.~N.~Nikolaev and B.~G.~Zakharov,
  J.\ Exp.\ Theor.\ Phys.\  {\bf 78}, 598 (1994)
  [Zh.\ Eksp.\ Teor.\ Fiz.\  {\bf 105}, 1117 (1994)].
  
\bibitem{Kopeliovich:1995an} 
  B.~Kopeliovich,
  in {\it Proceedings of the international workshop XXIII on Gross Properties 
  of Nuclei and Nuclear Excitations, Hirschegg, Austria, 1995}, edited by H.\ Feldmeyer and 
  W.\ N\"orenberg (Gesellschaft Schwerionenforschung, Darmstadt, 1995), p. 385, hep-ph/9609385.
  
\bibitem{Brodsky:1996nj} 
  S.~J.~Brodsky, A.~Hebecker and E.~Quack,
  Phys.\ Rev.\ D {\bf 55}, 2584 (1997).

\bibitem{Kopeliovich:1998nw} 
  B.~Z.~Kopeliovich, A.~V.~Tarasov and A.~Sch\"afer,
  Phys.\ Rev.\ C {\bf 59}, 1609 (1999)

\bibitem{Kopeliovich:2000fb} 
  B.~Z.~Kopeliovich, J.~Raufeisen and A.~V.~Tarasov,
  Phys.\ Lett.\ B {\bf 503}, 91 (2001).

\bibitem{Nikolaev:1990ja} 
  N.~N.~Nikolaev and B.~G.~Zakharov,
  Z.\ Phys.\ C {\bf 49}, 607 (1991).

\bibitem{bgbk} 
  J.~Bartels, K.~J.~Golec-Biernat and H.~Kowalski,
  Phys.\ Rev.\ D {\bf 66}, 014001 (2002).


\bibitem{KLP} 
  V.~G.~Kartvelishvili, A.~K.~Likhoded and V.~A.~Petrov,
  Phys.\ Lett.\  {\bf 78B}, 615 (1978).

\bibitem{KKSS} 
  B.~A.~Kniehl, G.~Kramer, I.~Schienbein and H.~Spiesberger,
  Eur.\ Phys.\ J.\ C {\bf 72}, 2082 (2012).

\bibitem{BKK} 
  J.~Binnewies, B.~A.~Kniehl and G.~Kramer,
  Phys.\ Rev.\ D {\bf 58}, 034016 (1998); \\
  B.~A.~Kniehl, G.~Kramer, I.~Schienbein and H.~Spiesberger,
  Eur.\ Phys.\ J.\ C {\bf 75}, no. 3, 140 (2015).

\bibitem{kst2} 
  B.~Z.~Kopeliovich, A.~Sch\"afer and A.~V.~Tarasov,
  Phys.\ Rev.\ D {\bf 62}, 054022 (2000).  
  
\bibitem{spots} 
  B.~Z.~Kopeliovich, I.~K.~Potashnikova, B.~Povh and I.~Schmidt,
  ``Evidences for two scales in hadrons,''
  Phys.\ Rev.\ D {\bf 76}, 094020 (2007).

\bibitem{gb} 
  J.~F.~Gunion and G.~Bertsch,
  Phys.\ Rev.\ D {\bf 25}, 746 (1982).


\bibitem{GBW}
  K.~J.~Golec-Biernat, M.~Wusthoff,
  Phys.\ Rev.\  D {\bf 59}, 014017 (1998).

\bibitem{iim} 
E. Iancu, K. Itakura, S. Munier, Phys. Lett. B {\bf 590}, 199  (2004).

\bibitem{kkt} 
D. Kharzeev, Y.V. Kovchegov and K. Tuchin, Phys. Lett. B {\bf  599}, 23  (2004).

\bibitem{dhj}
  A.~Dumitru, A.~Hayashigaki and J.~Jalilian-Marian,
  Nucl.\ Phys.\ A {\bf 765}, 464  (2006).

\bibitem{Goncalves:2006yt}
  V.~P.~Goncalves, M.~S.~Kugeratski, M.~V.~T.~Machado and F.~S.~Navarra,
  Phys.\ Lett.\  B {\bf 643}, 273  (2006).
  
\bibitem{buw} 
D. Boer, A. Utermann, E. Wessels,
Phys.\ Rev.\  D {\bf 77}, 054014 (2008).

\bibitem{kmw}
H.~Kowalski, L.~Motyka and G.~Watt,
  Phys.\ Rev.\ D {\bf 74}, 074016 (2006); \\
G.~Watt and H.~Kowalski,
  Phys.\ Rev.\  D {\bf 78}, 014016 (2008).

\bibitem{agbs} 
  J.~T.~de Santana Amaral, M.~B.~ Gay Ducati, M.~A.~Betemps, and G.~Soyez, 
  Phys.\ Rev.\ D {\bf 76}, 094018 (2007); \\
  E.~A.~F.~Basso, M.~B.~Gay Ducati and E.~G.~de Oliveira,
  Phys.\ Rev.\ D {\bf 87}, 074023 (2013).

\bibitem{Soyez2007}
G.~Soyez, Phys.\ Lett.\  B {\bf 655}, 32 (2007).

\bibitem{kt} 
H. Kowalski and D. Teaney, Phys. Rev. D {\bf 68}, 114005 (2003).

\bibitem{ipsatnewfit} 
A. H. Rezaeian, M. Siddikov, M. Van de Klundert and R. Venugopalan, Phys. Rev. D {\bf 87}, 034002 (2013).

\bibitem{amirs}
A. Rezaeian and I. Schmidt,  Phys.\ Rev.\  D {\bf 88}, 074016 (2013).


\bibitem{Blaettel:1993rd} 
  B.~Blaettel, G.~Baym, L.~L.~Frankfurt and M.~Strikman,
  Phys.\ Rev.\ Lett.\  {\bf 70}, 896 (1993).

\bibitem{Frankfurt:1993it} 
  L.~Frankfurt, G.~A.~Miller and M.~Strikman,
  Phys.\ Lett.\ B {\bf 304}, 1 (1993).

\bibitem{Frankfurt:1996ri} 
  L.~Frankfurt, A.~Radyushkin and M.~Strikman,
  Phys.\ Rev.\ D {\bf 55}, 98 (1997).

\bibitem{Nikolaev:1994cn} 
  N.~N.~Nikolaev and B.~G.~Zakharov,
  Phys.\ Lett.\ B {\bf 327}, 157 (1994).

\bibitem{KMR} 
  M.~A.~Kimber, A.~D.~Martin and M.~G.~Ryskin,
  Phys.\ Rev.\ D {\bf 63}, 114027 (2001).

\bibitem{Dmesons_lhcb} 
  R.~Aaij {\it et al.} [LHCb Collaboration],
  Nucl.\ Phys.\ B {\bf 871}, 1 (2013).

\bibitem{Bmesons_lhcb} 
  R.~Aaij {\it et al.} [LHCb Collaboration],
  JHEP {\bf 1308}, 117 (2013).

\bibitem{ct10} 
  H.~L.~Lai, M.~Guzzi, J.~Huston, Z.~Li, P.~M.~Nadolsky, J.~Pumplin and C.-P.~Yuan,
  Phys.\ Rev.\ D {\bf 82}, 074024 (2010).

\bibitem{Mangano:1991jk} 
  M.~L.~Mangano, P.~Nason and G.~Ridolfi,
  Nucl.\ Phys.\ B {\bf 373}, 295 (1992).

\bibitem{Hom:1976zn} 
  D.~C.~Hom {\it et al.},
  Phys.\ Rev.\ Lett.\  {\bf 37}, 1374 (1976).

\bibitem{Kaplan:1977kr} 
  D.~M.~Kaplan {\it et al.},
  Phys.\ Rev.\ Lett.\  {\bf 40}, 435 (1978).

\bibitem{Apanasevich:1997hm} 
  L.~Apanasevich {\it et al.} [Fermilab E706 Collaboration],
  Phys.\ Rev.\ Lett.\  {\bf 81}, 2642 (1998).

\bibitem{Apanasevich:1998ki} 
  L.~Apanasevich {\it et al.},
  Phys.\ Rev.\ D {\bf 59}, 074007 (1999).

\bibitem{Kopeliovich:2007yva} 
  B.~Z.~Kopeliovich, A.~H.~Rezaeian, H.~J.~Pirner and I.~Schmidt,
  Phys.\ Lett.\ B {\bf 653}, 210 (2007).

  
\bibitem{atlas_bjets} 
  G.~Aad {\it et al.} [ATLAS Collaboration],
  Eur.\ Phys.\ J.\ C {\bf 71}, 1846 (2011).

\bibitem{cms_bjets} 
  S.~Chatrchyan {\it et al.} [CMS Collaboration],
  JHEP {\bf 1204}, 084 (2012).

\end{thebibliography}
\end{document}